\documentclass[10pt,twocolumn,twoside,journal]{IEEEtran}
\usepackage{amsmath,graphicx,amssymb,mathtools,bm}
\usepackage{subfigure}
\usepackage{hyperref}

\usepackage{cite}
\ifCLASSINFOpdf
\else
\fi
\usepackage{algorithm}
\usepackage{algorithmic}
\usepackage{multirow}
\usepackage{xcolor}

\newtheorem{Theorem}{{Theorem}}
\newtheorem{Lemma}{{Lemma}}

\newcommand{\AD}{\mathrm {AD}}

\newcommand{\tr}{\mathrm {tr}}

\begin{document}
%\title{\Large{Performance Analysis of Massive MIMO Relay Systems With Variable Resolution ADCs/DACs Over Spatially Correlated Channels}} %\Large

\title{Spectral and Energy Efficiency of Multicell Massive MIMO With Variable-Resolution ADCs Over Correlated Rayleigh Fading Channels}


%\author{Author 1,
%        Author 2,
%        Author 3,
%        and Author 4 % <-this % stops a space

\author{Youzhi~Xiong,
        Sanshan~Sun, \textit{Member IEEE},
        Ning~Wei, \textit{Member IEEE},\\
        Li~Liu,
        and Zhongpei~Zhang, \textit{Member IEEE}
% <-this % stops a space
\thanks{This work was supported in part by National Natural  Science
Foundation of China (NSFC) under Grants 62101370, 61871070, 91938202, and 61831004, in part by Sichuan Science and Technology Program under Grant 2021YFG0013. (\textit{Corresponding author:
Youzhi Xiong}.)}
\thanks{Youzhi Xiong, Sanshan Sun, and Li Liu are with College of Physics and Electronic Engineering,
Sichuan Normal University, Chengdu, 610068, China. (e-mail: yzxiong@sicnu.edu.cn; sanshansun@hotmail.com; liuli@sicnu.edu.cn).

Ning Wei and Zhongpei Zhang are with National Key Laboratory of  Science and Technology  on Communications,
University of Electronic Science and Technology of China, Chengdu, 611731, China. (e-mail: wn@uestc.edu.cn; zhangzp@uestc.edu.cn)
}
}

\maketitle

\begin{abstract}
This paper analyzes the performance of multicell massive multiple-input and multiple-output (MIMO) systems with variable-resolution analog-to-digital converters (ADCs). In such an architecture, each ADC uses arbitrary quantization resolution to save power and hardware cost. Along this direction, we first introduce a quantization-aware channel estimator based on additive quantization noise model (AQNM) and linear minimum mean-squared error (LMMSE) estimate theory. Afterwards, by leveraging on the estimated channel state information (CSI), we derive the asymptotic expressions of achievable uplink spectral efficiency (SE) over spatially correlated Rayleigh fading channels for maximal ratio combining (MRC), quantization-aware multicell minimum mean-squared error (QA-M-MMSE) combining, and quantization-aware single-cell MMSE (QA-S-MMSE) combining, respectively. During the derivations, we consider the effect of quantization errors and resort to random matrix theory to achieve the asymptotic results. Finally, simulation results demonstrate that our theoretical analyses are correct and that the proposed quantization-aware estimator and combiners are more beneficial than the quantization-unaware counterparts. Besides, based on a generic power consumption model, it is shown that low-resolution ADCs can obtain the best tradeoff between SE and energy efficiency (EE) under multicell scenarios.
\end{abstract}

\begin{IEEEkeywords}
AQNM, spatially correlated Rayleigh fading, multicell massive MIMO, MMSE, variable-resolution ADCs.
\end{IEEEkeywords}

% \fi
\IEEEpeerreviewmaketitle

\section{Introduction}

\IEEEPARstart{T}{he} massive multiple-input and multiple-output (MIMO) paves the way to current and future wireless networks \cite{EJL2017,Marzetta:book}, e.g., the fifth-generation (5G) and beyond 5G wireless communication systems, by providing considerable spectral efficiency (SE). Although the massive MIMO can provide ubiquitous coverage and uniform service quality, significant power consumption and hardware cost, resulting from numerous radio frequency (RF) chains, become one of the major drawbacks in realizing pragmatic systems. To address this challenge in practical implementation, it was revealed in \cite{JLXYL2018} and \cite{JGAANB2020} that using low-resolution analog-to-digital converters (ADCs) is a feasible solution towards cost-and-energy efficient massive MIMO. In this context, quantization errors, caused by the low-resolution quantizers, are non-negligible. When it comes to multicell systems, both intra-cell and inter-cell signals will impact the quantization errors. In this regard, it is indispensable to analyze the performance under the circumstance of quantization errors.

\subsection{Prior Relevant Work}

Due to the fact that using low-resolution ADCs inevitably results in performance loss, it is crucial to evaluate the deleterious impact of quantization errors on system performance associated with various scenarios. Over the last few years, significant efforts have been dedicated to this topic. By using the maximal ratio combining (MRC) detection and additive quantization noise model (AQNM)/Bussgang decomposition,\footnote{Under the existence of quantization errors, the AQNM is an effective and simple approach to enable tractable analysis on performance. It has been shown in \cite{OE2021} that the AQNM is a special case of the Bussgang decomposition for distortion functions that satisfy a particular condition.} the achievable rate of single-cell massive MIMO systems with low-resolution ADCs was derived for Rayleigh channels in \cite{LSCH2015,SGMUC2017,LXSFY2019} and Rician fading channels in \cite{JLSZ2016,TJQJYZ2020,QYY2020,JJJRC2020}. It was found that performance loss can be compensated by increasing the number of antennas. To make a tradeoff between performance and power consumption, the collection of ADCs can be composed of 1-bit ADCs and a small part of high-resolution ADCs, which is called as the mixed-ADCs architecture. The corresponding performance analyses were studied in \cite{NW2016,JLZSX2017,QY2018II,HKBZ2020}. For multicell massive MIMO systems, low-resolution ADCs were considered in \cite{JWHGX2019,JYB2021}. To be specific, assuming that analog beamforming is used at the user side and MRC is employed at the base station (BS) side, Xu \textit{et al.} \cite{JWHGX2019} derived a lower bound for the achievable uplink rate over a non-cooperative multi-cell mmWave system. By contrast, considering a cooperative multi-cell massive MIMO systems, Choi \textit{et al.} \cite{JYB2021} investigated  coordinated multipoint (CoMP) beamforming and power control problems that consider the effect of quantization errors.

When it comes to relay networks, the authors of \cite{PHWX2017}, \cite{CACAZ2018} and \cite{JLZBO2019} derived the closed-form expressions of achievable rate when low-resolution ADCs, one-bit ADCs, and mixed ADCs are utilized at the relay, respectively. It was revealed that the performance loss caused by low resolution ADCs, e.g., 2-3 bits, is limited when the number of antennas at the relay is relatively large compared with the number of users. By integrating low-resolution ADCs into cell-free massive MIMO, recent papers\cite{XCXWHZ2019,YYMLH2020I,YMHLH2020II} analyzed the corresponding system performance. Specifically, asymptotic expression for each individual user was presented in \cite{XCXWHZ2019}. It was shown that the performance is mainly constrained by the ADC resolution at the user side under the condition that low-resolution ADCs are deployed at both the users and access points (APs). The authors of \cite{YYMLH2020I} derived the closed-form expression of achievable rate for MRC detection under a mixed-ADCs architecture. Over Rician fading channels, reference \cite{YMHLH2020II} provided an approximate uplink SE expression for MRC detection.

In practice, spatially correlated wireless channels are more accurate to characterize the propagation environments. In this regard, the authors in \cite{PHWGX2018} and \cite{QY2019} provided the performance analysis of single-cell massive MIMO in the presence of low-resolution ADCs and spatially correlated channels. By taking into account the spatial correlation and low-resolution ADCs used at the relay and BS, Dong \textit{et al.} in \cite{PHQG2020} derived the closed-form expression of achievable rate and discussed the power scaling law for the massive MIMO relay system. Considering that ADCs with arbitrary resolution profile are employed at the relay, Xiong \textit{et al.} \cite{YSNLZ2021} derived the closed-form expression of achievable rate over spatially correlated Rayleigh fading channels under both perfect and imperfect channel state information (CSI). This study also provided the condition under which low-resolution ADCs and receive spatial correlation dominate the performance loss.

\subsection{Contributions}

The most aforementioned works merely concentrate on single-cell systems and MRC scheme. As a matter of fact, it is essential to analyze the performance of MMSE-based receivers over multicell massive MIMO. Additionally, as mentioned in \cite{AJSTR2017,DEM2021,YSNLZ2021,SYM2020,XiongYZ2020}, it is of interest to consider a variable-resolution quantization, which can provides extra degrees-of-freedom for the design and optimization. Furthermore, in practical scenarios, spatial correlation calls for considerations during performance analysis. However, to the authors' best knowledge, such comprehensive investigation considering these factors is still missing in the literature. Thus, we will fill this gap in this article and the contributions are summarized as follows:
\begin{itemize}
  \item Focusing on multicell massive MIMO, we not only consider the inevitable spatial correlation, but also assume that all ADCs can use arbitrary resolution. In this context, we develop a quantization-aware channel estimator and make analytical analysis in the present of pilot contamination. In contrast to \cite{LSCH2015,SGMUC2017,LXSFY2019,JLSZ2016,TJQJYZ2020,QYY2020,JJJRC2020,NW2016,JLZSX2017,QY2018II,HKBZ2020,JWHGX2019,JYB2021,PHWX2017,CACAZ2018,JLZBO2019,XCXWHZ2019,YYMLH2020I,YMHLH2020II}, considering spatial correlation is particularly imperative for large arrays since practical channels are generally spatially correlated. Moreover, along our previous work \cite{YSNLZ2021}, we also consider variable-resolution ADC architecture, a more general case that includes 1-bit ADCs, pure low-resolution ADCs and mixed-ADC architectures discussed in the majority of existing works.
      %In principle, given an arbitrary resolution configuration of ADCs in multicell massive MIMO systems with spatially correlated CSI, one can obtain the corresponding lower bound of achievable SE by adopting the asymptotic results in this paper.
  \item Regarding performance analysis, we take into account the intra-cell interference, inter-cell interference, estimation errors, and quantization noise. Specifically, for the MRC, we consider the spatial correlation and use the exact variance of quantization noise instead of the approximation one used in \cite{PHWGX2018} and \cite{QY2019}. Additionally, many previous works in massive MIMO systems with MMSE combing only provide Monte Carlo simulations.\footnote{For instance, in \cite{EJL2017}, the SEs of MMSE-based combiners were simply obtained by averaging instantaneous SE via Monte Carlo methods rather than deriving closed-form ones. In \cite{ZJEB2021}, the authors stated that the exact closed-form expression of achievable rate cannot be obtained when using MMSE-based combining and provided Monte Carlo simulations as well.} Meanwhile, it is challenge to achieve the exact closed-form expressions of the achievable uplink SE for quantization-aware multicell minimum mean-squared error (QA-M-MMSE) and quantization-aware single-cell MMSE (QA-S-MMSE) combiners. Nevertheless, we resort to the AQNM, use-and-then-forget (UatF) technique, and random matrix theory to facilitate the derivation of the asymptotic alternatives over spatially correlated Rayleigh fading channels under imperfect CSI. Different from the existing literature \cite{LSCH2015,SGMUC2017,LXSFY2019,JLSZ2016,TJQJYZ2020,QYY2020,JJJRC2020,NW2016,JLZSX2017,QY2018II,HKBZ2020,JWHGX2019}, where only MRC over fixed resolution ADCs is considered, we provide the asymptotic analysis for quantization-aware MMSE-based combiners.
  \item Simulation results corroborate the correctness and accuracy of our asymptotic analyses. In taking into consideration the impact of quantization noise, it is demonstrated that the quantization-aware estimator and combiners are preferable than the quantization-unaware counterparts. Among the results, it can also be concluded that using low-resolution ADCs (e.g., 3-5 bits) in a multicell massive MIMO system is more advisable from the perspective of energy efficiency.
\end{itemize}

\subsection{Outline and Notations}

The remainder of the paper is organized as follows. In Section \ref{sec:sysmod}, we introduce the system model of multicell massive MIMO with variable-resolution ADCs. Quantization-aware channel estimator is developed in Section \ref{sec:channelEdtimation}. Under imperfect CSI, Section \ref{sec:AnalysisSEandEE} presents the spectral and energy efficiency analysis for both MRC and quantization-aware MMSE-based receivers. In Section \ref{sec:simResults}, simulation results and discussions are provided. This paper ends with a conclusion in Section \ref{sec:conclusion}.

Throughout this paper, $(\cdot)^\ast$, $(\cdot)^H$, and $(\cdot)^{T}$  represent the conjugate, Hermitian transpose, transpose, respectively. $\mathbf{a}_k$ denotes the $k$th column vector of $\mathbf{A}$, while $(\mathbf{A})_{ij}$ represents the $(i,j)$th element of $\mathbf{A}$. We use $\mathcal{CN}(\bm{\mu},\bm{\Sigma})$ to denote a complex Gaussian distribution with mean vector $\bm{\mu}$ and variance matrix $\bm{\Sigma}$. $\mathbb{E}\{\cdot\}$ and $\mathrm{tr}(\mathbf{A})$ are the expectation and trace of $\mathbf{A}$, respectively. $\mathrm{diag}\{a_1,\cdots,a_N\}$ and $\mathrm{diag}\{\mathbf{A}\}$ are both diagonal matrices constructed by $a_1,\cdots,a_N$ and the diagonal elements of $\mathbf{A}$ on the main diagonal, respectively.

\section{System Model}
\label{sec:sysmod}

%As depicted in Fig. \ref{fig:systemBiGAMPADC}, an uplink mmWave massive MIMO system with low-resolution ADCs is
%considered. In such a system, $N$ single-antenna users are simultaneously served by a base station (BS) deployed with uniform linear array (ULA) of $M$ antennas. Denote by $\Delta_r$ the normalized interval between any two adjacent receiver antenna, and by $L_n$ the number of physical paths between user $n$ and the BS. For the $l$th path, $\alpha_{l,n}$ and $\theta_{l,n}$ represent the path gain and

\begin{figure}[!t] %[htb]
\centering
  \includegraphics[width=3.0in]{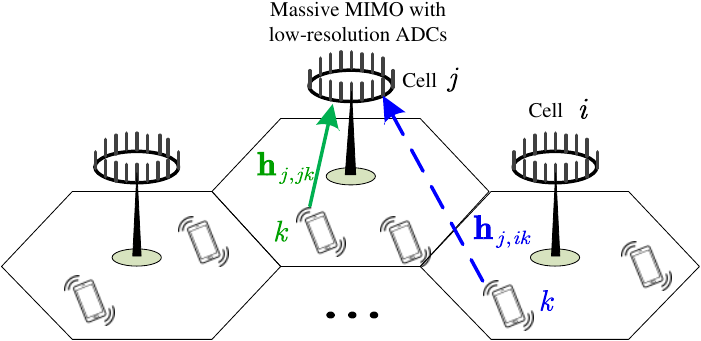}
%  \vspace{2.0cm}
%  \medskip
  \caption{Illustration of the multicell massive MIMO network with low-resolution ADCs at the BS. \label{fig:MaMIMOADC}}
\end{figure}

As depicted in Fig. \ref{fig:MaMIMOADC}, we consider a multicell and multiuser massive MIMO network with $L$ cells and $K$ single-antenna users per cell. Each cell is covered by a base station (BS) with $M$ antennas ($M\gg K$). The CSI between user $k$ in cell $i$ and all antennas of the BS in cell $j$ is denoted by $\mathbf{h}_{j,ik}\in \mathbb{C}^{M\times 1}$. In practice, wireless channels are spatially correlated. Therefore, in this study, we concentrate on spatially correlated Rayleigh fading channels, i.e.,
\begin{equation}\label{eq:correlatedCSI}
  \mathbf{h}_{j,ik}=\mathbf{R}_{j,ik}^{1/2}\mathbf{x}_{j,ik}.
\end{equation}
In (\ref{eq:correlatedCSI}), $\mathbf{x}_{j,ik}$ denotes the small-scale fading matrix with $\mathbf{x}_{j,ik}\thicksim \mathcal{CN}(\mathbf{0},\mathbf{I}_M)$, while $\mathbf{R}_{j,ik}$ denotes the spatial correlation matrix, which is assumed to be known at all BSs. The normalized trace $\beta_{j,ik}=\frac{1}{M}\mathrm{tr}(\mathbf{R}_{j,ik})$ determines the average channel gain, which can also be referred to as the large-scale fading coefficient. We assume that the BSs corresponding to all $L$ cells are equipped with low-resolution ADCs to reduce power consumption and hardware costs. In general, the propagation channels are assumed to be constant throughout one coherence time interval. Thus, time division multiplexing (TDD) is considered in this study.

\subsection{Uplink Pilot Transmission}

Suppose that $\bm{\phi}_{ik}\in\mathbb{C}^{\tau_p\times 1}$ denotes the pilot sequence used by user $k$ in cell $i$ and $\bm{\Phi}_i=[\bm{\phi}_{ik},\cdots,\bm{\phi}_{iK}]\in\mathbb{C}^{\tau_p\times K}$ represents the pilot matrix associated with cell $i$. It is assumed that $\bm{\phi}_{ik}$ has unit-magnitude elements to obtain a constant power level. Assuming that different users in a cell have orthogonal pilot sequences, it follows that $||\bm{\phi}_{ik}||^2=\tau_p$ and $\bm{\Phi}_i^H\bm{\Phi}_i=\tau_p\mathbf{I}_{\tau_p}$. During the stage of uplink pilot transmission, the received baseband signal at BS $j$ is expressed as
\begin{equation}\label{eq:receivedPilot}
  \mathbf{Y}^p_j=\mathbf{H}_{j,j}\mathbf{P}_{j}^{1/2}\bm{\Phi}^H_j+\sum_{i\neq j}^{L}\mathbf{H}_{j,i}\mathbf{P}_{i}^{1/2}\bm{\Phi}_i^H + \mathbf{N}^p_j,
\end{equation}
where $\mathbf{H}_{j,i}=[\mathbf{h}_{j,i1},\cdots,\mathbf{h}_{j,ik}]$ is the channel matrix between BS $j$ and all users in cell $i$ for $j=1,\cdots,L$, $i=1,\cdots,L$ and $k=1,\cdots,K$. In the right-hand side (RHS) of (\ref{eq:receivedPilot}), $\mathbf{P}_{i}^{1/2}=\mathrm{diag}(\sqrt{p_{i,1}},\cdots,\sqrt{p_{i,K}})$ denotes the transmit power matrix of the users in cell $i$ with $p_{i,k}$ being the transmit power of user $k$ in cell $i$, while $\mathbf{N}^{p}_j$ is the additive white Gaussian noise (AWGN) matrix, whose elements are independent and identically distributed (i.i.d.) and generated by $\mathcal{CN}(0,\sigma^2)$. Notice that $\sigma^2$ is the averaged power of the thermal noise.

It is assumed that the ADC pairs related to antenna $m$ at BS $j$ has $b_{j,m}$ quantization bits. Thus, $\mathbf{Y}^p_j$ is quantized by
\begin{equation}\label{eq:receivedPilotQuant0}
  \tilde{\mathbf{Y}}^p_j=\mathcal{Q}(\mathbf{Y}^p_j),
\end{equation}
where $\mathcal{Q}(\mathbf{Y}^p_j)$ is an element-wise quantization applied to the real and imaginary parts of $\mathbf{Y}^p_j$, respectively. Obeying to the majority of works \cite{SGMUC2017,LXSFY2019,JLSZ2016,TJQJYZ2020,QYY2020,JJJRC2020,NW2016,JLZSX2017,QY2018II,HKBZ2020,JWHGX2019,JYB2021,PHWX2017,CACAZ2018,JLZBO2019,XCXWHZ2019,YYMLH2020I,YMHLH2020II,PHWGX2018,QY2019,PHQG2020,YSNLZ2021}, we also adopt the AQNM to obtain a linearized approximation of the quantization $\mathcal{Q}(\mathbf{Y}^p_j)$ . Under the consideration of AQNM, the quantized version of $\mathbf{Y}^p_j$ can further be given by
\begin{equation}\label{eq:receivedPilotQuant}
  \tilde{\mathbf{Y}}^p_j=\mathcal{Q}(\mathbf{Y}^p_j)=\bm{\Sigma}^\AD_j\mathbf{Y}^p_j + \mathbf{Q}^p_j,
\end{equation}
where $\bm{\Sigma}^\AD_j=\mathrm{diag}(\alpha^\AD_{i,1},\cdots,\alpha^\AD_{i,M})$ denotes the distortion matrix determined by the ADCs at BS $j$. If the quantization bits of antenna $m$ at BS $j$ is $b^\AD_{j,m}$, the values of distortion factor $\alpha_{j,m}^{\AD}$ are exemplified in Table \ref{tab:normStep} for $b_{j,m}^{\AD}\leqslant 5$. When $b_{j,m}^{\AD}>5$, $\alpha_{m,n}^{\AD}$ is approximated as $\alpha_{j,m}^{\AD}\approx 1-\frac{\pi\sqrt{3}}{2}2^{-2b_{j,m}^{\AD}}$. Moreover, $\mathbf{Q}^p_j$ in (\ref{eq:receivedPilotQuant}) denotes the quantization noise which is uncorrelated with $\mathbf{Y}^p_j$ and each column of $\mathbf{Q}^p_j$ is assumed to follow the complex Gaussian distribution with zero mean and variance of
\begin{equation}\label{eq:quantizationCovarianceTraining}
\begin{split}
   &\mathbf{R}_{\mathbf{q}^p_j}  =\frac{1}{\tau_p}\bm{\Sigma}^\AD_j(\mathbf{I}_M-\bm{\Sigma}^\AD_j)\mathrm{diag}(\mathbb{E}\{\mathbf{Y}^p_j(\mathbf{Y}^p_j)^H\}) \\
     &= \bm{\Sigma}^\AD_j(\mathbf{I}_M-\bm{\Sigma}^\AD_j)\mathrm{diag}\left(\sum_{i=1}^{L}\mathbf{H}_{j,i}\mathbf{P}_{i}\mathbf{H}_{j,i}^H+\sigma^2\mathbf{I}_M\right).
\end{split}
\end{equation}

\begin{table}[!t]
  \centering
  \caption{Values of $\alpha_{j,m}^{\AD}$ for $b_{j,m}^{\AD}\leq 5$\cite{XiongYZ2020,YSNLZ2021}.\label{tab:normStep}}
  \begin{tabular}{|c|c|c|c|c|c|}
  \hline
  $b_{j,m}^{\AD}$ & 1 & 2 & 3 & 4 & 5\\
  \hline
  $\alpha_{j,m}^{\AD}$ & 0.6366 & 0.8825 & 0.96546 & 0.990503 & 0.997501\\
  \hline
  %$b$ & 5 & 6 & 7 & 8\\
%  \hline
%  $\Delta_{\mathrm{norm}}$ & 0.1881 & 0.1041 & 0.0569 & 0.0308\\
%  \hline
\end{tabular}
\end{table}

\subsection{Uplink Data Transmission}

During this phase, all users in each cell intend to transmit their uplink signal to the corresponding BS. Suppose that $\mathbf{s}_i$ is the symbols vector of the users in cell $i$ and that $\mathbf{s}_i$ has zero mean and unit variance for all $i=1,\cdots,L$. Then the received signal at BS $j$ is expressed as
\begin{equation}\label{eq:receivedData}
  \mathbf{y}_j=\mathbf{H}_{j,j}\mathbf{P}_{j}^{1/2}\mathbf{s}_j+\sum_{i\neq j}^{L}\mathbf{H}_{j,i}\mathbf{P}_{i}^{1/2}\mathbf{s}_i + \mathbf{n}_j,
\end{equation}
where $\mathbf{n}_j$, which follows $\mathcal{CN}(\mathbf{0},\sigma^2\mathbf{I}_M)$, is the AWGN vector at BS $j$. The second term of the RHS in (\ref{eq:receivedData}) represents the interference from the users in the other cells.

Suppose that the same ADC architecture is utilized for both uplink pilot and data transmissions. In that case, similarly to the stage of uplink pilot transmission, we also resort to the AQNM and thus obtain the quantized version of $\mathbf{y}_j$ via
\begin{equation}\label{eq:receivedDataQuant}
  \tilde{\mathbf{y}}_j=\mathcal{Q}(\mathbf{y}_j)=\bm{\Sigma}^\AD_j\mathbf{y}_j + \mathbf{q}_j,
\end{equation}
where $\mathbf{q}_j$ is the quantization noise which is uncorrelated with $\mathbf{y}_j$. Note that $\mathbf{q}_j$ is also assumed to follow the complex Gaussian distribution with zero mean and variance of
\begin{equation}\label{eq:quantizationCovarianceData}
\begin{split}
   &\mathbf{R}_{\mathbf{q}_j}  =\bm{\Sigma}^\AD_j(\mathbf{I}_M-\bm{\Sigma}^\AD_j)\mathrm{diag}(\mathbb{E}\{\mathbf{y}_j(\mathbf{y}_j)^H\}) \\
     &= \bm{\Sigma}^\AD_j(\mathbf{I}_M-\bm{\Sigma}^\AD_j)\mathrm{diag}\left(\sum_{i=1}^{L}\mathbf{H}_{j,i}\mathbf{P}_{i}\mathbf{H}_{j,i}^H+\sigma^2\mathbf{I}_M\right).
\end{split}
\end{equation}

\section{Uplink Channel Estimation}
\label{sec:channelEdtimation}

Since channel estimation is a prerequisite for coherent detection, in this section, we will develop a quantization-aware estimator to obtain CSI. In practical systems, symbols reserved for uplink training in one coherence block are inadequate under large $L$ and $K$ so one pilot set might be reused in different cells. We use $\mathcal{P}_{j,k}$ to denote the set of all users that utilize the same pilot sequence as user $k$ in cell $j$. On this basis, $\mathcal{P}_{j,k}$ is mathematically defined as
\begin{equation}\label{eq:setPjk}
  \mathcal{P}_{j,k} = \{(i,k^\prime): \bm{\phi}_{i,k^\prime}=\bm{\phi}_{j,k}, i=1,\cdots,L,k^\prime=1,\cdots,K\}.
\end{equation}

Suppose that BS $j$ intends to estimate the channel $\mathbf{h}_{j,ik}$ from an arbitrary user $k$ in cell $i$. \footnote{When $i=j$, BS $j$ will estimate the channels of its own users. } The BS can correlate $\tilde{\mathbf{Y}}^p_j$ with the pilot sequence $\bm{\phi}_{i,k}$ associated with this user. As a result, the processed received pilot signal is $\tilde{\mathbf{y}}^p_{j,ik}=\tilde{\mathbf{Y}}^p_j\bm{\phi}_{i,k}$, which is expanded into
\begin{equation}\label{eq:sinalCorrelate}
\begin{split}
   \tilde{\mathbf{y}}^p_{j,ik} & =\sqrt{p_{i,k}}\tau_p\bm{\Sigma}^\AD_j\mathbf{h}_{j,ik} + \sum_{(l,k^\prime)\in\mathcal{P}_{i,k}\backslash(i,k)}\sqrt{p_{l,k^\prime}}\tau_p\bm{\Sigma}^\AD_j\mathbf{h}_{j,lk^\prime} \\
     &+\bm{\Sigma}^\AD_j\mathbf{N}^p_j\bm{\phi}_{i,k} + \mathbf{Q}^p_j\bm{\phi}_{i,k}.
\end{split}
\end{equation}
The second term of the RHS in (\ref{eq:sinalCorrelate}) denotes the pilot contamination produced by the users in $\mathcal{P}_{j,k}$ except user $k$ in cell $j$. Besides, in (\ref{eq:sinalCorrelate}), it holds that $\mathbf{N}^p_j\bm{\phi}_{i,k}\thicksim \mathcal{CN}(\mathbf{0},\sigma^2\tau_p\mathbf{I}_M)$ and $\mathbf{Q}^p_j\bm{\phi}_{i,k}\thicksim \mathcal{CN}(\mathbf{0},\tau_p\mathbf{R}_{\mathbf{q}^p_j})$.

In what follows, we will investigate the estimation process for $\mathbf{h}_{j,ik}$ based on $\tilde{\mathbf{y}}^p_{j,ik}$. By referring to \cite{Kay:book} and taking the impact of quantization into account, the quantization-aware MMSE estimate of $\mathbf{h}_{j,ik}$ is given by
\begin{equation}\label{eq:channelEst}
  \hat{\mathbf{h}}_{j,ik}=\sqrt{p_{i,k}}\mathbf{R}_{j,ik}\bm{\Sigma}^\AD_j\bm{\Psi}_{j,ik}\tilde{\mathbf{y}}^p_{j,ik},
\end{equation}
where
\begin{equation}\label{eq:Psijik}
  \bm{\Psi}_{j,ik} = \left(\sum_{(l,k^\prime)\in\mathcal{P}_{i,k}\backslash(i,k)}p_{l,k^\prime}\tau_p\bm{\Sigma}^\AD_j\mathbf{R}_{j,lk^\prime}\bm{\Sigma}^\AD_j
  +\mathbf{Z}_{j}^p\right)^{-1}
\end{equation}
with $\mathbf{Z}_{j}^p=\sigma^2(\bm{\Sigma}^\AD_j)^2+\bar{\mathbf{R}}_{\mathbf{q}^p_j}$. It is worthwhile to mention that
\begin{equation}\label{eq:quantizationCovarianceAverage}
  \bar{\mathbf{R}}_{\mathbf{q}^p_j}=\bm{\Sigma}^\AD_j(\mathbf{I}_M-\bm{\Sigma}^\AD_j)\mathrm{diag}\left(\sum_{i=1}^{L}\sum_{k=1}^{K}p_{i,k}\mathbf{R}_{j,ik}+\sigma^2\mathbf{I}_M\right)
\end{equation}
is the approximation of ${\mathbf{R}}_{\mathbf{q}^p_j}$. We here use $\bar{\mathbf{R}}_{\mathbf{q}^p_j}$ rather than ${\mathbf{R}}_{\mathbf{q}^p_j}$ because the perfect CSI in (\ref{eq:quantizationCovarianceTraining}) is unavailable during the stage of channel estimation and only the spatial correlation matrix $\mathbf{R}_{j,ik}$ is assumed to be known at the BSs.

According to the linear MMSE estimate theory in \cite{Kay:book}, the estimation error is modeled as $\tilde{\mathbf{h}}_{j,ik}={\mathbf{h}}_{j,ik}-\hat{\mathbf{h}}_{j,ik}$. Moreover, the variance matrices of $\hat{\mathbf{h}}_{j,ik}$ and $\tilde{\mathbf{h}}_{j,ik}$ are, respectively, calculated as
\begin{equation}\label{eq:covarianceBjik}
  \mathbf{B}_{j,ik}=\mathbb{E}\{\hat{\mathbf{h}}_{j,ik}\hat{\mathbf{h}}^H_{j,ik}\}=p_{i,k}\tau_p\mathbf{R}_{j,ik}\bm{\Sigma}^\AD_j\bm{\Psi}_{j,ik}\bm{\Sigma}^\AD_j\mathbf{R}_{j,ik},
\end{equation}
and
\begin{equation}\label{eq:covarianceCjik}
  \mathbf{C}_{j,ik}=\mathbb{E}\{\tilde{\mathbf{h}}_{j,ik}\tilde{\mathbf{h}}^H_{j,ik}\}=\mathbf{R}_{j,ik}-\mathbf{B}_{j,ik}.
\end{equation}
The estimation quality is evaluated by the normalized MSE, which is given by
\begin{equation}\label{eq:NMSE}
  \mathsf{NMSE}_{j,ik}=\frac{\mathrm{tr}(\mathbf{C}_{j,ik})}{\mathrm{tr}(\mathbf{R}_{j,ik})}.
\end{equation}
If user $k$ in cell $i$ uses the same pilot sequence as user $k^\prime$ in cell $i^\prime$ and $\mathbf{R}_{j,ik}$ is invertible, then we have
\begin{equation}\label{eq:CSIcorr}
  \hat{\mathbf{h}}_{j,i^\prime k^\prime}=\frac{\sqrt{p_{i^\prime,k^\prime}}}{\sqrt{p_{i,k}}}\mathbf{R}_{j,i^\prime k^\prime}(\mathbf{R}_{j,ik})^{-1}\hat{\mathbf{h}}_{j,i k}.
\end{equation}
This implies that $\hat{\mathbf{h}}_{j,i^\prime k^\prime}$ and $\hat{\mathbf{h}}_{j,i k}$ are correlated with each other and leads to
\begin{equation}\label{eq:corrh1h2}
  \mathbb{E}\{\hat{\mathbf{h}}_{j,i^\prime k^\prime}\hat{\mathbf{h}}^H_{j,i k}\} =\frac{\sqrt{p_{i^\prime,k^\prime}}}{\sqrt{p_{i, k}}}\mathbf{R}_{j,i^\prime k^\prime}(\mathbf{R}_{j,ik})^{-1}\mathbf{B}_{j,i k},
\end{equation}
which will be used in the sequel.

\section{Analysis on Spectral and Energy Efficiency}
\label{sec:AnalysisSEandEE}

In this section, we investigate the achievable SE corresponding to the uplink data transmission under different linear receive combiners. Without loss of generality, we merely focus on the SE derivations for user $k$ in cell $j$. Suppose that BS $j$ selects the combining vector $\mathbf{v}_{j,k}$, depending on the estimated channels obtained from the uplink pilot transmission, to recover the desired information of its $k$th user. By doing so, BS $j$ correlates the quantized signal $\tilde{\mathbf{y}}_j$ in (\ref{eq:receivedDataQuant}) to obtain $\hat{s}_{j,k}=\mathbf{v}_{j,k}^H(\bm{\Sigma}^\AD_j)^{-1}\tilde{\mathbf{y}}_j$, which can be unfolded into
\begin{equation}\label{eq:correlateQuantizedData}
\begin{split}
     \hat{s}_{j,k}&=  \underbrace{\sqrt{p_{j,k}}\mathbb{E}\{\mathbf{v}_{j,k}^H\hat{\mathbf{h}}_{j,jk}^Hs_{j,k}\}}_{\mathrm{Desired\,signal}} \\
     &+\underbrace{\sqrt{p_{j,k}}(\mathbf{v}_{j,k}^H\hat{\mathbf{h}}_{j,jk}^Hs_{j,k}-\mathbb{E}\{\mathbf{v}_{j,k}^H\hat{\mathbf{h}}_{j,jk}^Hs_{j,k}\})}_{\mathrm{Uncertainty\,of \,desired\,signal}}\\
     &+ \underbrace{\sum_{k^\prime\neq k}^{K}\sqrt{p_{j,k^\prime}}\mathbf{v}_{j,k}^H\hat{\mathbf{h}}_{j,jk^\prime}^Hs_{j,k^\prime}}_{\mathrm{Intra-cell\, interference}}\\
     &+\underbrace{\sum_{i\neq j}^{L}\sum_{k^\prime=1}^{K}\sqrt{p_{i,k^\prime}}\mathbf{v}_{j,k}^H\hat{\mathbf{h}}_{j,ik^\prime}^Hs_{i,k^\prime}}_{\mathrm{Inter-cell\, interference}}\\
     &+\underbrace{\sum_{i=1}^{L}\sum_{k^\prime=1}^{K}\sqrt{p_{i,k^\prime}}\mathbf{v}_{j,k}^H\tilde{\mathbf{h}}_{j,ik^\prime}^Hs_{i,k^\prime}}_{\mathrm{Interference\, from\, estimation\,error}}\\
       &+ \underbrace{\mathbf{v}_{j,k}^H\mathbf{n}_j}_{\mathrm{Noise}} +\underbrace{\mathbf{v}_{j,k}^H(\bm{\Sigma}^\AD_j)^{-1}\mathbf{q}_j}_{\mathrm{Quantization \,noise}}.
  \end{split}
\end{equation}
In the RHS of (\ref{eq:correlateQuantizedData}), the term $\sqrt{p_{j,k}}\mathbb{E}\{\mathbf{v}_{j,k}^H\hat{\mathbf{h}}_{j,jk}^Hs_{j,k}\}$ \footnote{We do not use the instantaneous $\sqrt{p_{j,k}}\mathbf{v}_{j,k}^H\hat{\mathbf{h}}_{j,jk}^Hs_{j,k}$ but only its statistics $\sqrt{p_{j,k}}\mathbb{E}\{\mathbf{v}_{j,k}^H\hat{\mathbf{h}}_{j,jk}^Hs_{j,k}\}$. The main reason is that the statistics is easy to compute given the statistical property and can be easily acquired over a long-time scale in practical systems.} is treated as true desired signal. The other terms of the RHS in (\ref{eq:correlateQuantizedData}) can be regarded as \textit{effective noise}. According to the use-and-then-forget (UatF) bound \footnote{As approved in \cite{QSKHM2014}, $\log_2(1+\mathbb{E}\{X\}/\mathbb{E}\{Y\})$ is a lower bound of $\mathbb{E}\{\log
_2(1+X/Y)\}$. Moreover, when $X$ and $Y$ are both the sums of the nonnegative random variables and $M\rightarrow\infty$, the common approximation $\mathbb{E}\{\log
_2(1+X/Y)\}\thickapprox \log_2(1+\mathbb{E}\{X\}/\mathbb{E}\{Y\})$ is tight.} mentioned in \cite{EJL2017} and \cite{Marzetta:book}, the achievable uplink SE (a.k.a the achievable uplink rate) of user $k$ in cell $j$ is lower bounded by
\begin{equation}\label{eq:SEUatF}
  \mathsf{SE}_{j,k}=\frac{\tau_u}{\tau_c}\log_2(1+\frac{A_{j,k}}{B_{j,k}+C_{j,k}+D_{j,k}+E_{j,k}+F_{j,k}+G_{j,k}}),
\end{equation}
in which
\begin{equation}\label{eq:ExpeAjk}
  A_{j,k}=p_{j,k}|\mathbb{E}\{\mathbf{v}_{j,k}^H\hat{\mathbf{h}}_{j,jk}\}|^2,
\end{equation}
\begin{equation}\label{eq:ExpeBjk}
  B_{j,k}=p_{j,k}\mathbb{E}\{|\mathbf{v}_{j,k}^H\hat{\mathbf{h}}_{j,jk}|^2\}-A_{j,k},
\end{equation}
\begin{equation}\label{eq:ExpeCjk}
  C_{j,k}=\sum_{k^\prime\neq k}^{K}p_{j,k^\prime}\mathbb{E}\{|\mathbf{v}_{j,k}^H\hat{\mathbf{h}}_{j,jk^\prime}|^2\}
\end{equation}
\begin{equation}\label{eq:ExpeDjk}
  D_{j,k}=\sum_{i\neq j}^{L}\sum_{k^\prime=1}^{K}p_{i,k^\prime}\mathbb{E}\{|\mathbf{v}_{j,k}^H\hat{\mathbf{h}}_{j,ik^\prime}|^2\},
\end{equation}
\begin{equation}\label{eq:ExpeEjk}
  E_{j,k}=\sum_{i=1}^{L}\sum_{k^\prime=1}^{K}p_{i,k^\prime}\mathbb{E}\{|\mathbf{v}_{j,k}^H\tilde{\mathbf{h}}_{j,ik^\prime}|^2\},
\end{equation}
\begin{equation}\label{eq:ExpeFjk}
  F_{j,k}=\sigma^2\mathbb{E}\{||\mathbf{v}_{j,k}||^2\},
\end{equation}
\begin{equation}\label{eq:ExpeGjk}
  G_{j,k}=\mathbb{E}\{\mathbf{v}_{j,k}^H(\bm{\Sigma}^\AD_j)^{-1}\mathbf{R}_{\mathbf{q}_j}(\bm{\Sigma}^\AD_j)^{-1}\mathbf{v}_{j,k}\}.
\end{equation}
All the expectations are with respect to the channel realizations. In what follows, we will derive the asymptotic SEs for three linear combiners.

\subsection{Asymptotic Spectral Efficiency of MRC}

If MR combining is used, it follows that $\mathbf{v}_{j,k}=\hat{\mathbf{h}}_{j,jk}$. Based on the MMSE estimator in Section \ref{sec:channelEdtimation}, we have the following Theorem \ref{lemma:SEMRC}.
\begin{Theorem}\label{lemma:SEMRC}
For MR combining, the terms within the logarithm of (\ref{eq:SEUatF}) can be obtained in closed form. In particular, $A_{j,k}$ and $F_{j,k}$ are given by (\ref{eq:AjkMRC}) and (\ref{eq:FjkMRC}), respectively. $B_{j,k}+C_{j,k}+D_{j,k}+E_{j,k}$ and $G_{j,k}$ are, respectively, given by (\ref{eq:BjkEjikMRC}) and (\ref{eq:GjkMRC}) at the top of next page.
\begin{equation}\label{eq:AjkMRC}
  A_{j,k}=p_{j,k}\left|\mathrm{tr}(\mathbf{B}_{j,jk})\right|^2,
\end{equation}
\begin{figure*}[!t]
\begin{equation}\label{eq:BjkEjikMRC}
B_{j,k}+C_{j,k}+D_{j,k}+E_{j,k}=\sum_{i=1}^{L}\sum_{k^\prime=1}^{K} p_{i,k^\prime}(\mathrm{tr}(\mathbf{B}_{j,jk}\mathbf{R}_{j,ik^\prime})
     +\sum_{(i,k^\prime)\in\mathcal{P}_{j,k}}p^2_{i,k^\prime}p_{j,k}\tau_p^2|\mathrm{tr}(\mathbf{R}_{j,ik^\prime}\bm{\Sigma}^\AD_j\bm{\Psi}_{j,jk}\bm{\Sigma}^\AD_j\mathbf{R}_{j,jk})|^2-A_{j,k},
\end{equation}
\begin{equation}\label{eq:GjkMRC}
\begin{split}
   G_{j,k} & = \sum_{(i,k^\prime)\in\mathcal{P}_{j,k}} \sum_{m=1}^{M}\frac{1-\alpha^\AD_{j,m}}{\alpha^\AD_{j,m}}\bigg(
   \frac{p^2_{i,k^\prime}}{p_{j,k}}(\bm{\Xi})_{mm}+p_{i,k^\prime}(\mathbf{C}_{j,ik^\prime})_{mm}(\mathbf{B}_{j,jk})_{mm}\bigg)\\
     & +\sum_{(i,k^\prime)\notin\mathcal{P}_{j,k}}\sum_{m=1}^{M}\frac{1-\alpha^\AD_{j,m}}{\alpha^\AD_{j,m}}\bigg(
     p_{i,k^\prime}(\mathbf{R}_{j,ik^\prime})_{mm}(\mathbf{B}_{j,jk})_{mm}\bigg)+\sigma^2\mathrm{tr}(\mathbf{B}_{j,jk}((\bm{\Sigma}^\AD_j)^{-1}-\mathbf{I}_M)),
\end{split}
\end{equation}
\hrulefill
\vspace*{4pt}
\end{figure*}
\begin{equation}\label{eq:FjkMRC}
  F_{j,k}=\sigma^2\mathrm{tr}(\mathbf{B}_{j,jk}).
\end{equation}
Note in (\ref{eq:GjkMRC}) that $\bm{\Xi}$ is given by
\begin{equation}\label{eq:Xi}
  \bm{\Xi}=\mathbf{R}_{j,ik^\prime}(\mathbf{R}_{j,jk})^{-1}\mathbf{V}^m_{j,jk}(\mathbf{R}_{j,jk})^{-1}\mathbf{R}_{j,ik^\prime}.
\end{equation}
Moreover, the $(m_1,m_2)$th component of $\mathbf{V}^m_{j,jk}$ in (\ref{eq:Xi}) is given by
\begin{equation}\label{eq:Vmjjk}
\begin{split}
    & (\mathbf{V}^m_{j,jk})_{m_1m_2} \\
     & =\sum_{\bar{m},\bar{\bar{m}}}\left(t_{m_1\bar{m}}t^\ast_{m_2\bar{m}}|t_{m\bar{\bar{m}}}|^2
     +t_{m_1\bar{m}}t^\ast_{m\bar{m}}t^\ast_{m_2\bar{\bar{m}}}t_{m\bar{\bar{m}}}\right),
\end{split}
\end{equation}
in which $t_{pq}$ is the $(p,q)$th component of $\mathbf{T}=\mathbf{B}_{j,jk}^{1/2}$.
\end{Theorem}
\begin{IEEEproof}
The proof is available in Appendix \ref{app:lemmaMRC}.
\end{IEEEproof}

Substituting (\ref{eq:AjkMRC}) to (\ref{eq:GjkMRC}) back into (\ref{eq:SEUatF}) yields the asymptotic closed-form SE for the MR combining.

\subsection{Asymptotic Spectral Efficiency of Quantization-Aware Multicell MMSE}

If BS $j$ intends to mitigate interference originating from the users in cell $j$ except user $k$ and all users in the other cells and to alleviate the impact of quantization errors, we should design a quantization-aware multicell minimum mean-squared error (QA-M-MMSE) combining $\mathbf{v}_{j,k}^{\mathrm{M}}$. To this end, when only the estimated CSI are available, it follows that
\begin{equation}\label{eq:vjkMMMSE}
  \mathbf{v}_{j,k}^{\mathrm{M}}=p_{j,k}\left(\sum_{i=1}^{L}\hat{\mathbf{H}}_{j,i}\mathbf{P}_{i}(\hat{\mathbf{H}}_{j,i})^H+\mathbf{Z}_j^\mathrm{M}+\sigma^2\mathbf{I}_M\right)^{-1}\hat{\mathbf{h}}_{j,jk},
\end{equation}
in which $\hat{\mathbf{H}}_{j,i}=[\hat{\mathbf{h}}_{j,i1},\cdots,\hat{\mathbf{h}}_{j,ik}]$ and $\mathbf{Z}_j^\mathrm{M}=\sum_{i=1}^{L}\sum_{k^\prime=1}^{K}p_{i,k^\prime}\mathbf{C}_{j,ik^\prime}+(\bm{\Sigma}^\AD_j)^{-1}\bar{\mathbf{R}}_{\mathbf{q}_j}(\bm{\Sigma}^\AD_j)^{-1}$. It should be pointed out that
\begin{equation}\label{eq:quantizationCovarianceAverageData}
  \bar{\mathbf{R}}_{\mathbf{q}_j}=\bm{\Sigma}^\AD_j(\mathbf{I}_M-\bm{\Sigma}^\AD_j)\mathrm{diag}\left(\sum_{i=1}^{L}\sum_{k=1}^{K}p_{i,k}\mathbf{R}_{j,ik}+\sigma^2\mathbf{I}_M\right)
\end{equation}
is approximately equal to $\mathbf{R}_{\mathbf{q}_j}$. In comparison with \cite{EJL2017}, it is obvious that the combining vector in (\ref{eq:vjkMMMSE}) takes into account the effect of quantization errors, i.e., $\bar{\mathbf{R}}_{\mathbf{q}_j}$. Based on the MMSE estimator in Section \ref{sec:channelEdtimation}, we have the following Theorem \ref{lemma:SEMMMSE}, which is not provided in \cite{EJL2017}.
\begin{Theorem}\label{lemma:SEMMMSE}
For the quantization-aware multicell MMSE combining, the terms within the logarithm of (\ref{eq:SEUatF}) can be obtained in closed form as follows.
\begin{equation}\label{eq:AjkMMMSE}
  A_{j,k}\rightarrow p_{j,k}\left|\frac{\frac{\mathrm{tr}(\mathbf{B}_{j,jk}\mathbf{T}_{j,jk})}{M}}{1+\frac{\mathrm{tr}(\mathbf{B}_{j,jk}\mathbf{T}_{j,jk})}{M}}\right|^2,
\end{equation}
\begin{equation}\label{eq:BjkMMMSE}
  B_{j,k}\rightarrow  0,
\end{equation}
\begin{equation}\label{eq:CjkMMMSE}
  C_{j,k}\rightarrow  \sum_{k^\prime\neq k}^{K} \frac{p_{j,k^\prime}\frac{\mathrm{tr}(\mathbf{B}_{j,jk}\mathbf{T}^{\prime}_{j,jk^\prime})}{M^2}}{\left|1+\frac{\mathrm{tr}(\mathbf{B}_{j,jk}\mathbf{T}_{j,jk})}{M}\right|^2\left|1+\frac{\mathrm{tr}(\mathbf{B}_{j,jk^\prime}\mathbf{T}_{j,jk^\prime})}{M}\right|^2},
\end{equation}
\begin{equation}\label{eq:DjkMMMSE}
  D_{j,k}\rightarrow  \sum_{i\neq j}^{K}\sum_{k^\prime=1}^{K} \frac{p_{j,k^\prime}\frac{\mathrm{tr}(\mathbf{B}_{j,jk}\mathbf{T}^{\prime}_{j,ik^\prime})}{M^2}}{\left|1+\frac{\mathrm{tr}(\mathbf{B}_{j,jk}\mathbf{T}_{j,jk})}{M}\right|^2\left|1+\frac{\mathrm{tr}(\mathbf{B}_{j,ik^\prime}\mathbf{T}_{j,ik^\prime})}{M}\right|^2},
\end{equation}
\begin{equation}\label{eq:EjkMMMSE}
  E_{j,k}\rightarrow \sum_{i=1}^{K}\sum_{k^\prime=1}^{K} \frac{p_{j,k^\prime}\frac{\mathrm{tr}(\mathbf{B}_{j,jk}\tilde{\mathbf{T}}^{\prime}_{j,ik^\prime})}{M^2}}{\left|1+\frac{\mathrm{tr}(\mathbf{B}_{j,jk}\mathbf{T}_{j,jk})}{M}\right|^2},
\end{equation}
\begin{equation}\label{eq:FjkMMMSE}
  F_{j,k}\rightarrow \frac{\sigma^2\frac{\mathrm{tr}(\mathbf{B}_{j,jk}\mathbf{T}^{\prime}_{jk,\mathrm{n}})}{M^2}}{\left|1+\frac{\mathrm{tr}(\mathbf{B}_{j,jk}\mathbf{T}_{j,jk})}{M}\right|^2},
\end{equation}
\begin{equation}\label{eq:GjkMMMSE}
  G_{j,k}\rightarrow \frac{\frac{\mathrm{tr}(\mathbf{B}_{j,jk}\mathbf{T}^{\prime}_{jk,\mathrm{q}})}{M^2}}{\left|1+\frac{\mathrm{tr}(\mathbf{B}_{j,jk}\mathbf{T}_{j,jk})}{M}\right|^2}.
\end{equation}
Moreover, $\mathbf{T}_{j,jk}$ is obtained via (\ref{eq:Gamma}) in Lemma \ref{lemma:AB} by letting $\mathbf{A}=\mathbf{B}_{j,jk}$, $\bm{\Delta}_{ik^\prime}=\frac{p_{ik^\prime}}{p_{jk}}\mathbf{B}_{j,ik^\prime}$, $\mathbf{D}=\frac{\mathbf{Z}^{\mathrm{M}}_{j}}{p_{jk}M}$, and $\alpha=\frac{\sigma^2}{p_{jk}M}$. Subsequently, $\mathbf{T}^{\prime}_{j,ik^\prime}$, $\tilde{\mathbf{T}}^{\prime}_{j,ik^\prime}$, $\mathbf{T}^{\prime}_{jk,\mathrm{n}}$, and $\mathbf{T}^{\prime}_{jk,\mathrm{q}}$ are obtained via (\ref{eq:GammaPrime}) in Lemma \ref{lemma:ABC} by letting $\mathbf{C}=\mathbf{B}_{j,ik^\prime}$, $\mathbf{C}=\mathbf{C}_{j,ik^\prime}$, $\mathbf{C}=\mathbf{I}_{M}$, and $\mathbf{C}=(\bm{\Sigma}^\AD_j)^{-1}\bar{\mathbf{R}}_{\mathbf{q}_j}(\bm{\Sigma}^\AD_j)^{-1}$, respectively.
\end{Theorem}
\begin{IEEEproof}
The proof is available in Appendix \ref{app:lemmaMMMSE}.
\end{IEEEproof}

Plugging (\ref{eq:AjkMMMSE}) to (\ref{eq:GjkMMMSE}) into (\ref{eq:SEUatF}) gives the asymptotic closed-form SE for the quantization-aware M-MMSE combining.

\subsection{Asymptotic Spectral Efficiency of Quantization-Aware Single-cell MMSE}

Although the quantization-aware multicell MMSE combining can, to a certain extent, suppress the interference from other interfering users, this scheme needs to estimate the CSI between the BS in a considered cell and the users in other cells. This might give rise to high computational complexity and complicated implementation. If BS $j$ only estimates the channels relating to its own users, we develop a quantization-aware single-cell minimum mean-squared error (QA-S-MMSE) combining $\mathbf{v}_{j,k}^{\mathrm{S}}$, which is expressed as
\begin{equation}\label{eq:vjkSMMSE}
  \mathbf{v}_{j,k}^{\mathrm{S}}=p_{j,k}\left(\hat{\mathbf{H}}_{j,j}\mathbf{P}_{j}(\hat{\mathbf{H}}_{j,j})^H+\mathbf{Z}_j^\mathrm{S}+\sigma^2\mathbf{I}_M\right)^{-1}\hat{\mathbf{h}}_{j,jk},
\end{equation}
where $\mathbf{Z}_j^\mathrm{S}=\sum_{i\neq j}^{L}\sum_{k^\prime=1}^{K}p_{i,k^\prime}\mathbf{R}_{j,ik^\prime}+\sum_{k^\prime=1}^{K}p_{j,k^\prime}\mathbf{C}_{j,jk^\prime}+(\bm{\Sigma}^\AD_j)^{-1}\bar{\mathbf{R}}_{\mathbf{q}_j}(\bm{\Sigma}^\AD_j)^{-1}$.
Compared with \cite{EJL2017}, the combining vector $\mathbf{v}_{j,k}^{\mathrm{S}}$ in (\ref{eq:vjkSMMSE}) also takes into consideration the effect of quantization errors, i.e., $\bar{\mathbf{R}}_{\mathbf{q}_j}$. Based on the MMSE estimator in Section \ref{sec:channelEdtimation}, we have the following Theorem \ref{lemma:SESMMSE}, which is not provided in \cite{EJL2017}.
\begin{Theorem}\label{lemma:SESMMSE}
For the quantization-aware single-cell MMSE combining, the terms within the logarithm of (\ref{eq:SEUatF}) can be obtained in closed form as follows.
\begin{equation}\label{eq:AjkSMMSE}
  A_{j,k}\rightarrow p_{j,k}\left|\frac{\frac{\mathrm{tr}(\mathbf{B}_{j,jk}\bm{\Gamma}_{j,jk})}{M}}{1+\frac{\mathrm{tr}(\mathbf{B}_{j,jk}\bm{\Gamma}_{j,jk})}{M}}\right|^2,
\end{equation}
\begin{equation}\label{eq:BjkSMMSE}
  B_{j,k}\rightarrow  0,
\end{equation}
\begin{equation}\label{eq:CjkSMMSE}
  C_{j,k}\rightarrow  \sum_{k^\prime\neq k}^{K} \frac{p_{j,k^\prime}\frac{\mathrm{tr}(\mathbf{B}_{j,jk}\bm{\Gamma}^{\prime}_{j,jk^\prime})}{M^2}}{\left|1+\frac{\mathrm{tr}(\mathbf{B}_{j,jk}\bm{\Gamma}_{j,jk})}{M}\right|^2\left|1+\frac{\mathrm{tr}(\mathbf{B}_{j,jk^\prime}\bm{\Gamma}_{j,jk^\prime})}{M}\right|^2},
\end{equation}
\begin{equation}\label{eq:DjkSMMSE}
\begin{split}
   D_{j,k} &\rightarrow \sum_{\substack{ i\neq j\\(i,k^\prime)\in\mathcal{P}_{j,k}}}^{K} p_{i,k^\prime} \left|\frac{\frac{\mathrm{tr}\left(\frac{\sqrt{p_{i k^\prime}}}{\sqrt{p_{j k}}}\mathbf{R}_{j,i k^\prime}(\mathbf{R}_{j,jk})^{-1}\mathbf{B}_{j,j k}\bm{\Gamma}_{j,jk}\right)}{M}}
   {1+\frac{\mathrm{tr}(\mathbf{B}_{j,jk}\bm{\Gamma}_{j,jk})}{M}}\right|^2 \\
     & +\sum_{\substack{ i\neq j\\(i,k^\prime)\notin\mathcal{P}_{j,k}}}^{K} p_{i,k^\prime} \frac{\frac{\mathrm{tr}(\mathbf{B}_{j,jk}\bm{\Gamma}^{\prime}_{j,ik^\prime})}{M^2}}
     {\left|1+\frac{\mathrm{tr}(\mathbf{B}_{j,jk}\bm{\Gamma}_{j,jk})}{M}\right|^2},
\end{split}
\end{equation}
\begin{equation}\label{eq:EjkSMMSE}
  E_{j,k}\rightarrow \sum_{i=1}^{K}\sum_{k^\prime=1}^{K} \frac{p_{j,k^\prime}\frac{\mathrm{tr}(\mathbf{B}_{j,jk}\tilde{\bm{\Gamma}}^{\prime}_{j,ik^\prime})}{M^2}}{\left|1+\frac{\mathrm{tr}(\mathbf{B}_{j,jk}\bm{\Gamma}_{j,jk})}{M}\right|^2},
\end{equation}
\begin{equation}\label{eq:FjkSMMSE}
  F_{j,k}\rightarrow \frac{\sigma^2\frac{\mathrm{tr}(\mathbf{B}_{j,jk}\bm{\Gamma}^{\prime}_{jk,\mathrm{n}})}{M^2}}{\left|1+\frac{\mathrm{tr}(\mathbf{B}_{j,jk}\bm{\Gamma}_{j,jk})}{M}\right|^2},
\end{equation}
\begin{equation}\label{eq:GjkSMMSE}
  G_{j,k}\rightarrow \frac{\frac{\mathrm{tr}(\mathbf{B}_{j,jk}\bm{\Gamma}^{\prime}_{jk,\mathrm{q}})}{M^2}}{\left|1+\frac{\mathrm{tr}(\mathbf{B}_{j,jk}\bm{\Gamma}_{j,jk})}{M}\right|^2}.
\end{equation}
Moreover, $\bm{\Gamma}_{j,jk}$ is obtained via (\ref{eq:Gamma}) in Lemma \ref{lemma:AB} by letting $L=1$, $\mathbf{A}=\mathbf{B}_{j,jk}$, $\bm{\Delta}_{jk^\prime}=\frac{p_{jk^\prime}}{p_{jk}}\mathbf{B}_{j,jk^\prime}$, $\mathbf{D}=\frac{\mathbf{Z}^{\mathrm{S}}_{j}}{p_{jk}M}$, and $\alpha=\frac{\sigma^2}{p_{jk}M}$. Next, $\bm{\Gamma}^{\prime}_{j,ik^\prime}$, $\tilde{\bm{\Gamma}}^{\prime}_{j,ik^\prime}$, $\bm{\Gamma}^{\prime}_{jk,\mathrm{n}}$, and $\bm{\Gamma}^{\prime}_{jk,\mathrm{q}}$ are obtained via (\ref{eq:GammaPrime}) in Lemma \ref{lemma:ABC} by letting $\mathbf{C}=\mathbf{B}_{j,ik^\prime}$, $\mathbf{C}=\mathbf{C}_{j,ik^\prime}$, $\mathbf{C}=\mathbf{I}_{M}$, and $\mathbf{C}=(\bm{\Sigma}^\AD_j)^{-1}\bar{\mathbf{R}}_{\mathbf{q}_j}(\bm{\Sigma}^\AD_j)^{-1}$, respectively.
\end{Theorem}
\begin{IEEEproof}
The proof is available in Appendix \ref{app:lemmaSMMSE}.
\end{IEEEproof}

By substitution of (\ref{eq:AjkSMMSE}) to (\ref{eq:GjkSMMSE}) back into (\ref{eq:SEUatF}),  we get the asymptotic closed-form SE for the quantization-aware S-MMSE combining.

\subsection{Energy Efficiency}

\begin{figure}[!t] %[htb]
\centering
  \includegraphics[width=3.0in]{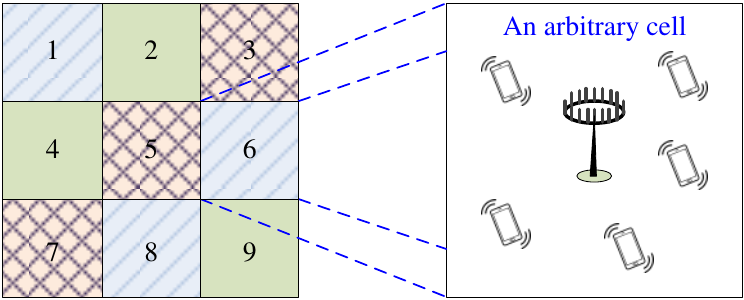}
%  \vspace{2.0cm}
%  \medskip
  \caption{Illustration of the setup with 9 cells located on a square of $3\times 3$ grid. The cells with same color are use the same pilot.\label{fig:pilotAssignment}}
\end{figure}

According to the power consumption model in \cite{JLZBO2019,YZNBY2019,YSNLZ2021}, the energy efficiency is defined  mathematically as
\begin{equation}\label{eq:EEmodel}
  \eta_{\mathrm{EE}}=\frac{W\sum_{j=1}^{L}\sum_{k=1}^{K}\mathsf{SE}_{j,k}}{P_{\mathrm{total}}} \mathrm{bits/Joule},
\end{equation}
where $P_{\mathrm{total}}$ denotes the total power consumption concerning all BSs in a multicell massive MIMO, $\mathsf{SE}_{j,k}$ is given by (\ref{eq:SEUatF}), and $W$ indicates the transmission bandwidth with $W=20$ MHz.
The power consumption of an ADC with $b^\AD_{j,m}$ quantization bits is
\begin{equation}\label{eq:powerConsumptionADC}
  P_{\mathrm{ADC}}=\frac{3V_{dd}^2L_{\mathrm{min}}(2W+f_{cor})}{10^{-0.1525b^\AD_{j,m}+4.838}},
\end{equation}
where $V_{dd}=3$ is the power supply of a converter, $L_{\mathrm{min}}=0.5\times10^{-6}$ represents the minimum channel length under the given CMOS technology, $f_{cor}=10^{6}$ denotes the corner frequency of the $1/f$ noise \cite{JLZBO2019}.
In this context, $P_{\mathrm{total}}$ can be calculated as
\begin{equation}\label{eq:totalPower}
\begin{split}
   P_{\mathrm{total}}& =LM(2P_{\mathrm{mix}}+P_{\mathrm{filt}}+P_{\mathrm{filr}}+P_{\mathrm{LNA}}+P_{\mathrm{IFA}})\\
   &+2LP_{\mathrm{syn}}+\sum_{j=1}^{L}\sum_{m=1}^{M}2(c^\AD_{j,m}P_{\mathrm{AGC}}+P_{\mathrm{ADC}}),
\end{split}
\end{equation}
where $P_{\mathrm{mix}}$, $P_{\mathrm{filt}}$, $P_{\mathrm{filr}}$, $P_{\mathrm{LNA}}$, $P_{\mathrm{IFA}}$, $P_{\mathrm{syn}}$, $P_{\mathrm{AGC}}$ denote the power consumption pertaining to the mixer, the filters at the transmitter, the filters at the receiver, low noise amplifiers (LNA), the intermediate frequency amplifier (IFA), the frequency synthesizer, and the automatic gain control (AGC), respectively. In addition, $c^\AD_{j,m}$ is given by
\begin{equation*}
  c^\AD_{j,m}=
  \begin{cases}
    0, & b^\AD_{j,m}=1 \\
    1, & b^\AD_{j,m}>1
  \end{cases},
\end{equation*}
which is an indicator describing whether or not antenna $m$ at BS $j$ uses 1-bit ADC pair.

\section{Numerical Results and Discussions}
\label{sec:simResults}

In the simulation to follow, we use a 9-cell setup, as depicted in Fig. \ref{fig:pilotAssignment}. Each cell covers a square with area $0.25\mathrm{km}\times 0.25\mathrm{km}$ and is deployed on a square of $3\times 3$ cells. The large-scale fading coefficient $\beta_{j,ik}$ is modeled as $\beta_{j,ik}=-148.1\mathrm{dB}-37.6\alpha\log_{10}(d_{j,ik}/1\mathrm{km})$, where $d_{j,ik}$ is the distance between BS $j$ and user $k$ in cell $i$. $5$ users are independently and uniformly distributed in each cell, at the center of which there exists a BS with $M=30$ antennas. We consider communication over $20$ MHz bandwidth and the noise power is $-94$ dBm, including noise figure with $7$ dB. Suppose that the local scattering model in \cite{EJL2017} is utilized and antenna arrays have half-wavelength spacing. If multipaths arrive from the far-field of the arrays, the $(m,n)$th entry of spatial correlation matrix $\mathbf{R}_{j,ik}$ can be computed as \cite{EJL2017}
\begin{equation}\label{eq:correlationmatrix}
  [\mathbf{R}_{j,ik}]_{mn}=\beta_{j,ik}\int e^{\mathrm{j}\pi(m-n)\sin(\psi)}f(\bar{\varphi})\mathrm{d}\bar{\varphi},
\end{equation}
where $\bar{\varphi}$ denotes the angle of a multipath component. Notice that $\bar{\varphi}={\varphi}+\delta$ with $\varphi$ being a deterministic nominal angle and $\delta$ being a random deviation from $\varphi$. It is obvious from (\ref{eq:correlationmatrix}) that $\mathbf{R}_{j,ik}$ is a Toeplitz matrix. In the simulation, Gaussian distributed
deviation $\delta\thicksim \mathcal{N}(0,\sigma^2_{\varphi})$ is used. The standard deviation $\sigma_{\varphi}\geqslant0$ is called the angular standard deviation (ASD). Unless stated otherwise, we consider $\sigma_{\varphi}=10^\circ$, $\tau_c=200$ samples and $\tau_p=fK$ samples with $f$ being the pilot reuse factor associated with 9-cell setup. In what follows, we let $f=3$ and the corresponding pilot assignment is shown in Fig. \ref{fig:pilotAssignment}. The Monte Carlo simulation results are obtained by averaging instantaneous SE over $100$ realizations of small-scale fading channels and $100$ realizations of the random locations of users, while the asymptotic results are obtained by averaging asymptotic SEs in Theorem \ref{lemma:SEMRC}, Theorem \ref{lemma:SEMMMSE}, and Theorem \ref{lemma:SESMMSE} over $100$ realizations of the random locations of users, respectively.

\begin{figure}[!t] %[htb]
\centering
  \includegraphics[width=2.7in]{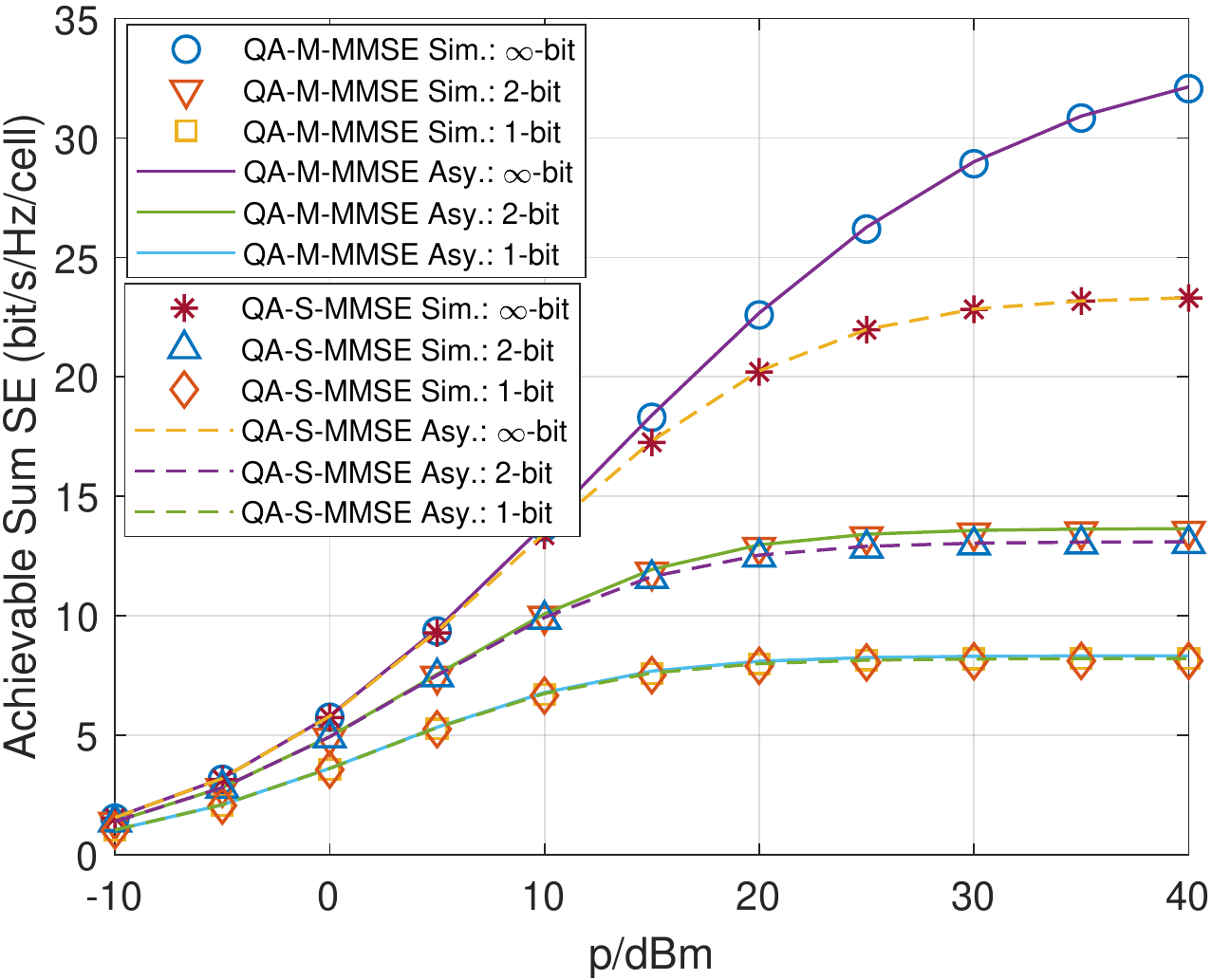}
%  \vspace{2.0cm}
%  \medskip
  \caption{Comparison of achievable SE between simulated results and asymptotic results for QA-M-MMSE and QA-S-MMSE. $L=9$,  $M=30$, $K=5$, $f=3$, and $\sigma_\varphi=10^\circ$.\label{fig:SEvsPowerMMSE}}
\end{figure}

\begin{figure}[!t] %[htb]
\centering
  \includegraphics[width=2.7in]{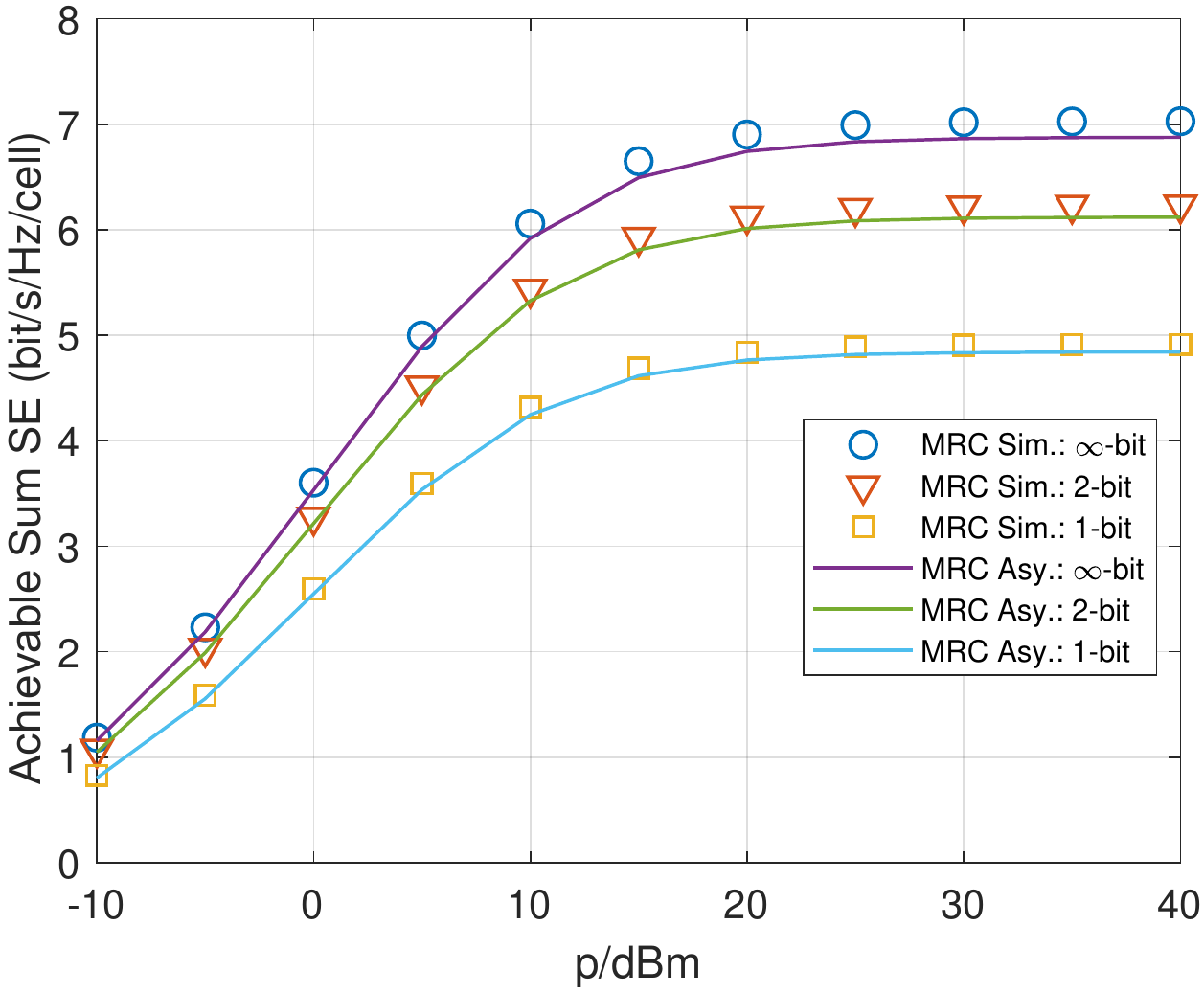}
%  \vspace{2.0cm}
%  \medskip
  \caption{Comparison of achievable SE between simulated results and asymptotic results for MRC. $L=9$, $M=30$, $K=5$, $f=3$, and $\sigma_\varphi=10^\circ$.\label{fig:SEvsPowerMRC}}
\end{figure}

Under different quantization bits, Fig. \ref{fig:SEvsPowerMMSE} and Fig. \ref{fig:SEvsPowerMRC} compare the asymptotic results and the simulated ones for MMSE-based and MRC combiners, respectively. Note that ``Asy." and ``Sim." represent the asymptotic and simulation results, respectively. It is evident from Fig. \ref{fig:SEvsPowerMMSE} that the asymptotic results of QA-M-MMSE and QA-S-MMSE match with the corresponding simulation results in a high degree of accuracy, especially for the case without quantization. Meanwhile, for the MRC, it is observed from Fig. \ref{fig:SEvsPowerMRC} that the asymptotic results substantially coincide with the simulation ones and the gap is inconspicuous. These observations approve the tightness of our asymptotic analyses relating to Theorem \ref{lemma:SEMRC}, Theorem \ref{lemma:SEMMMSE}, and Theorem \ref{lemma:SESMMSE}. Moreover, from the results in Fig. \ref{fig:SEvsPowerMMSE}, we also observe that the performance of QA-M-MMSE is superior to that of QA-S-MMSE in the case of un-quantization. The main reason is that the QA-M-MMSE can make a trade-off of suppression between intra-cell and inter-cell interference. Instead, the QA-S-MMSE can only mitigate intra-cell interference. However, at the regime of low-resolution ADCs, the performance gap between QA-M-MMSE and QA-S-MMSE shrinks as the quantization bit decreases. The reason is that the interference caused by quantization errors, compared with that produced by other users, is the dominant factor resulting in performance loss, notably under 1-bit quantization.

\begin{figure}[!t] %[htb]
\centering
  \includegraphics[width=2.7in]{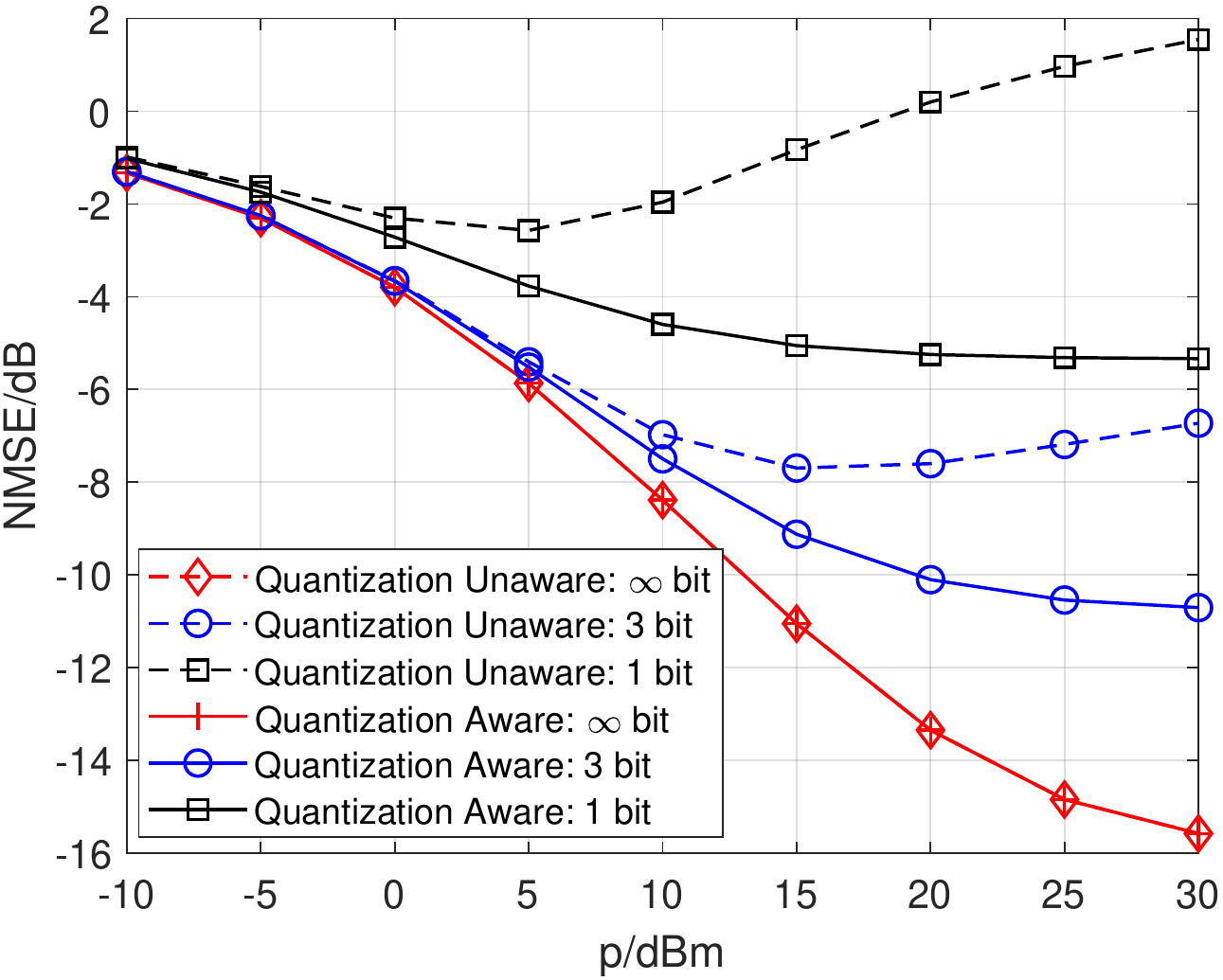}
%  \vspace{2.0cm}
%  \medskip
  \caption{Performance of channel estimation for quantization-aware and quantization-unaware estimators under different quantization bits. $L=9$, $M=30$, $K=5$, $f=3$, and $\sigma_\varphi=10^\circ$.\label{fig:NMSEvsPowerAwareUnaware}}
\end{figure}

Fig. \ref{fig:NMSEvsPowerAwareUnaware} exemplifies the estimation quality of quantization-aware and quantization-unaware MMSE estimators in terms of NMSE. It is noticeable that considering the impact of quantization errors can improve the performance of channel estimation, particularly in the high SNR region under low-resolution quantization. The main reason is that the quantization-aware estimator takes into account the variance of quantization noise during channel estimation, as shown in (\ref{eq:channelEst}) and (\ref{eq:Psijik}), whereas the quantization-unaware one does not. In addition, we can also observe that increasing transmit power cannot completely compensate the impact of low resolution quantization. When there exist quantization errors, the NMSE approaches to an asymptotic value as we augment the transmit power. This saturation occurs because the interference caused by quantization errors, proportional to the transmit power, dominate the loss of NMSE compared with the thermal noise.

\begin{figure}[!t] %[htb]
\centering
  \includegraphics[width=2.7in]{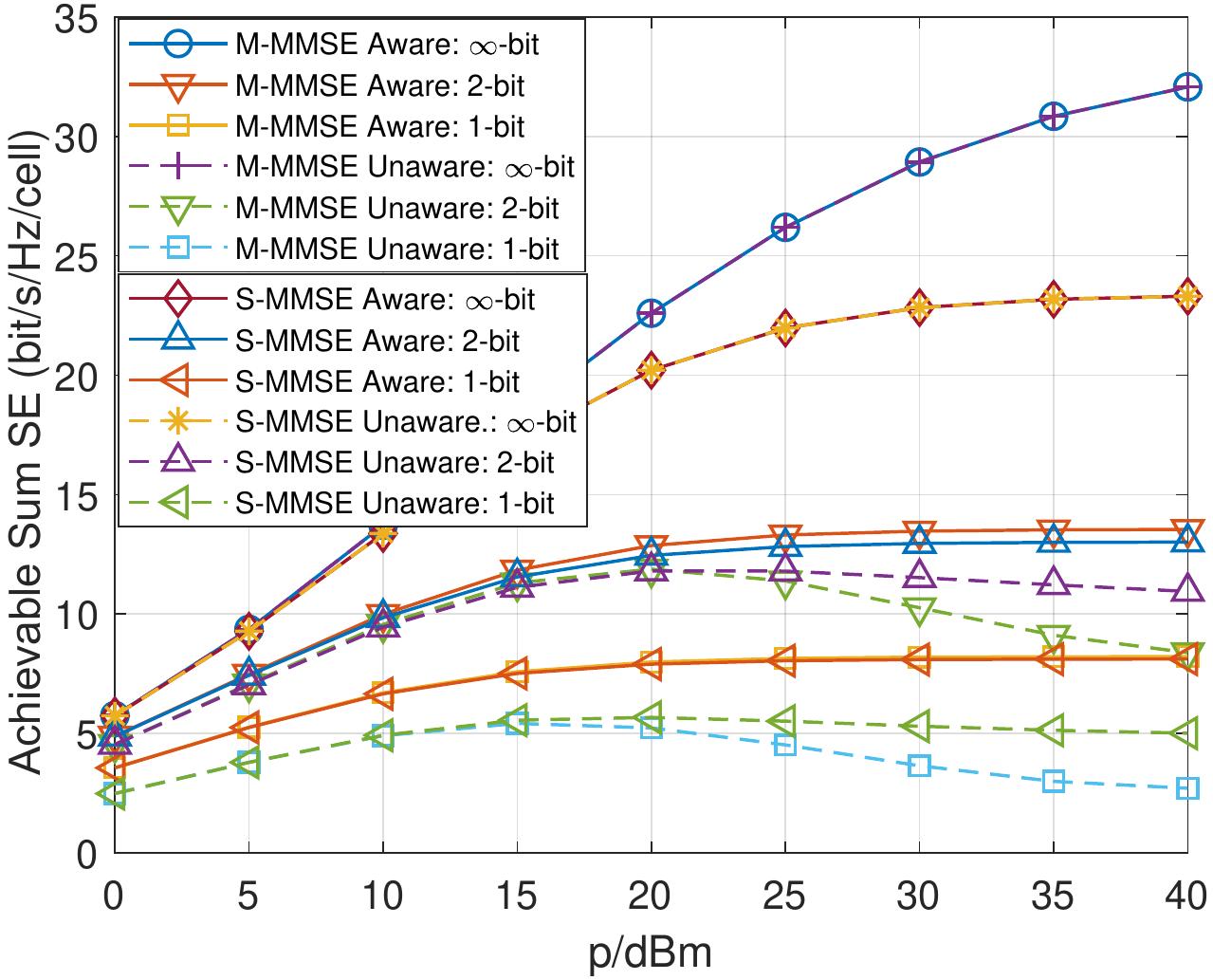}
%  \vspace{2.0cm}
%  \medskip
  \caption{Achievable SE for quantization-aware and quantization-unaware MMSE-based combiners under different quantization bits. $L=9$, $M=30$, $K=5$, $f=3$, and $\sigma_\varphi=10^\circ$.\label{fig:SEvsPowerAwareUnawareMMSE}}
\end{figure}

Fig. \ref{fig:SEvsPowerAwareUnawareMMSE} shows the achievable sum SEs for quantization-aware and quantization-unaware MMSE-based combiners, respectively. Again, we find that considering the impact of quantization errors can ameliorate system performance. The reason is that the proposed quantization-aware MMSE combining can be treated as a spatial whitening filter and is an optimal linear combiner in terms of typical mean-squared error (MSE) criterion. As a result, the impact of the distortion caused by quantization can be alleviated to a certain extent. Moreover, for quantization-unaware MMSE combiners, the S-MMSE is more beneficial than the M-MMSE. The main reason is that the interference caused by higher channel estimation errors dominate the performance loss. In this case, the M-MMSE, which aims to exploit the estimated channel associated with the users in other cells, cannot effectively compensate the interference from other cells under relatively inaccurate channel estimate degraded by quantization errors. Meanwhile, suppressing inter-cell interference is obtained at the cost of reducing the performance of mitigating intra-cell interference.

\begin{figure}[!t] %[htb]
\centering
  \includegraphics[width=2.7in]{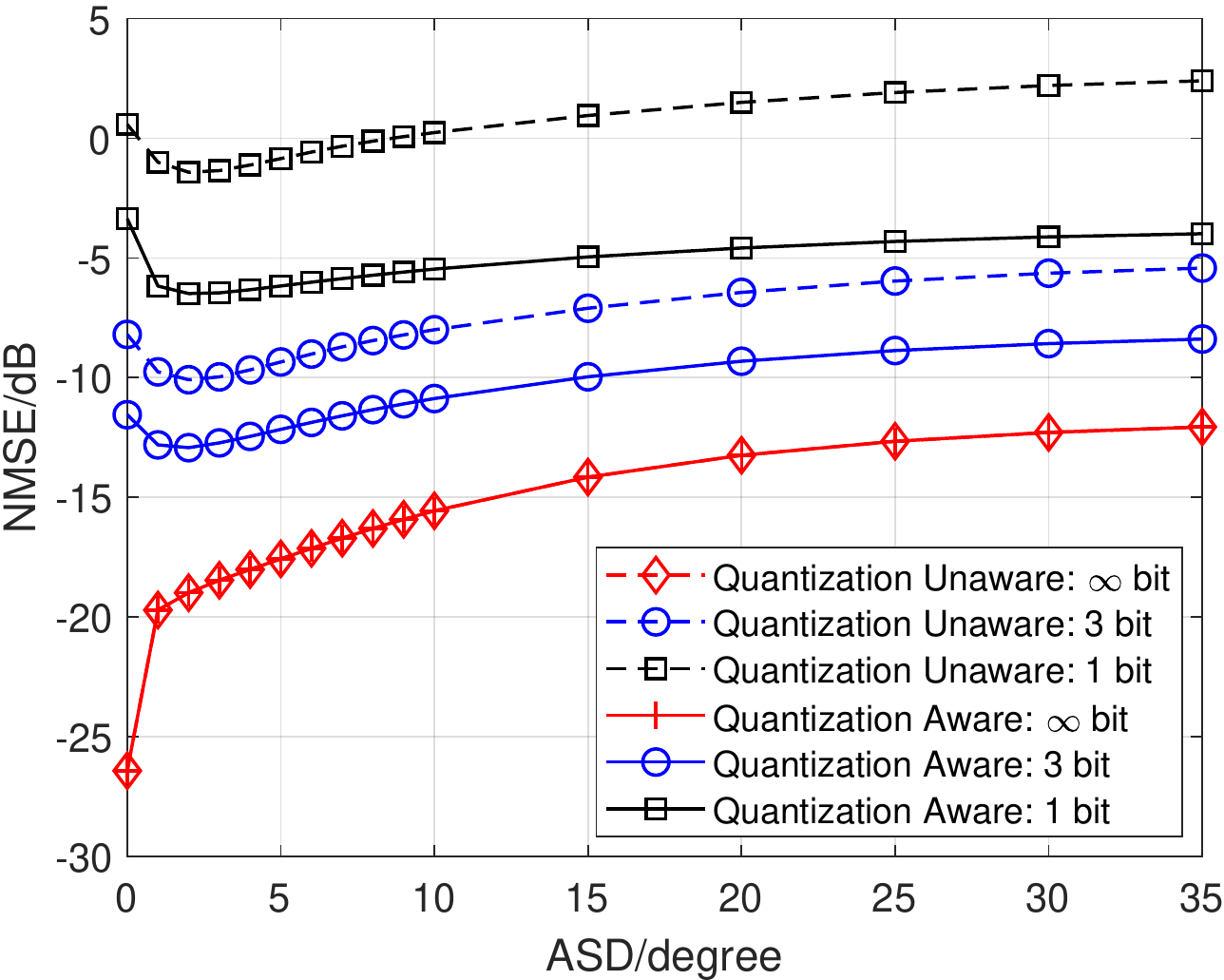}
%  \vspace{2.0cm}
%  \medskip
  \caption{Performance of channel estimation for quantization-aware and quantization-unaware estimators under different quantization bits. $L=9$, $M=30$, $K=5$, $f=3$, and $p=30$ dBm.\label{fig:NMSEvsASDAwareUnaware}}
\end{figure}

\begin{figure}[!t] %[htb]
\centering
  \includegraphics[width=2.7in]{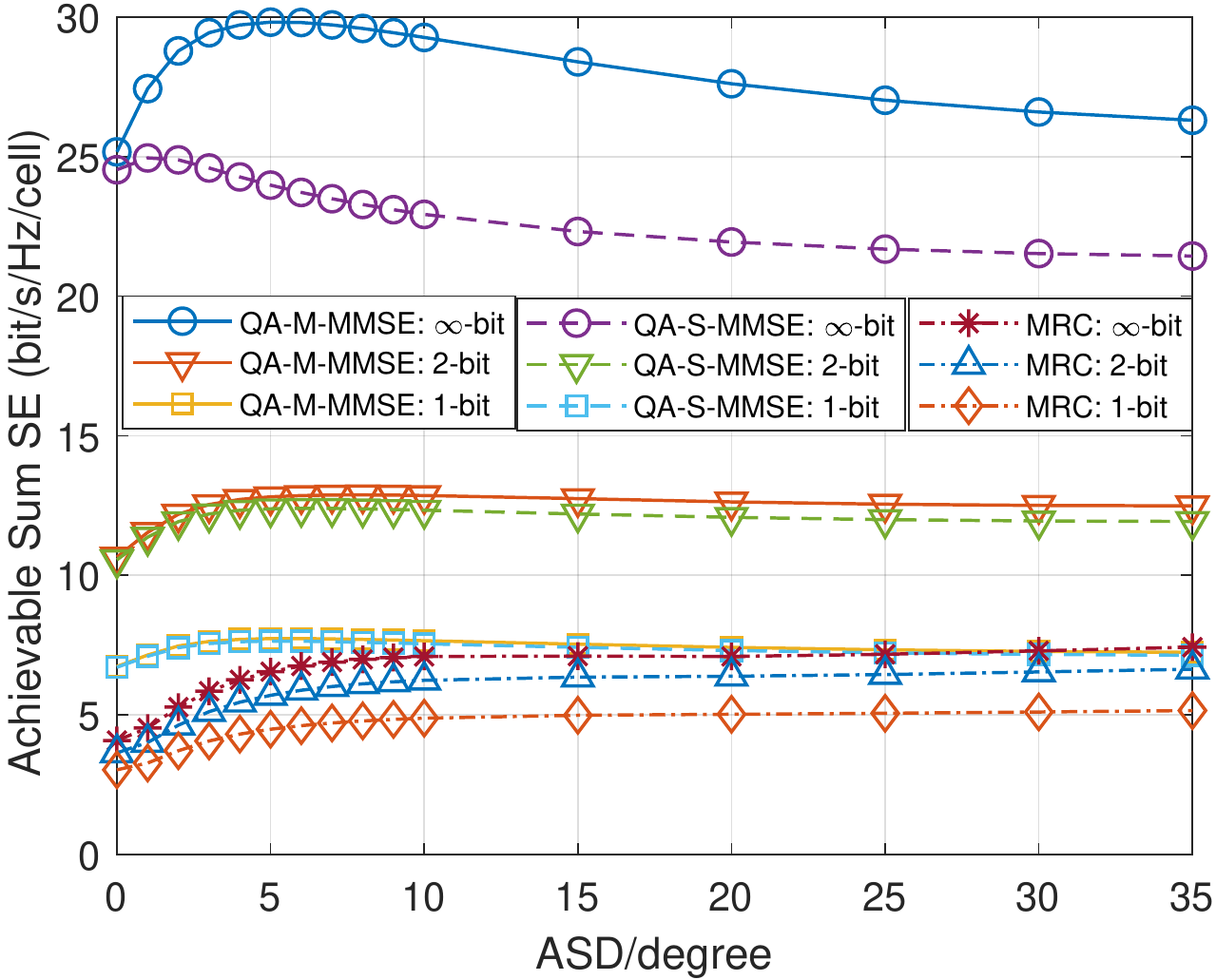}
%  \vspace{2.0cm}
%  \medskip
  \caption{Achievable SE for quantization-aware and quantization-unaware combiners based on MMSE along with MRC. $L=9$, $M=30$, $K=5$, $f=3$, and $p=30$ dBm.\label{fig:SEvsASDAwareMMSEMRC}}
\end{figure}

Fig. \ref{fig:NMSEvsASDAwareUnaware} assesses the performance of channel estimation in terms of NMSE for different channel correlation under the existence of pilot contamination. In the context of un-quantization, we observe that the NMSE reduces as the ASD decreases (i.e., towards higher spatial correlation) for both quantization-aware and quantization-unaware MMSE estimators. In fact, this two estimators are equivalent if there do not exist quantization errors. However, in the case of low-resolution quantization, the NMSE first decreases but then rises as the ASD changes to large degrees. The main reason is explained as follows. On the one hand, strong spatial correlation is beneficial due to the fact that most of the channel's variance lies in a few eigenvalues when ASD is small and that it is much harder to estimate weaken eigendirections than stronger ones. On the other hand, strong spatial correlation is detrimental because of the fact that the quantization errors are adversely affected by spatial correlation. This implies that there exists an ADS to achieve the optimal NMSE.\footnote{This optimal ASD can be found experimentally. In principle, to optimize the ADS is an interesting and challenging topic, which will be left as future work.}

Fig. \ref{fig:SEvsASDAwareMMSEMRC} illustrates the achievable sum SEs for different channel correlation under the existence of pilot contamination. We see that the performance is not so good when the spatial channel correlation is strong (with very small ASD). For MMSE-based combiners, the SE first increases but then reduces as the ASD changes to large values. For MRC, the SE increases as the ADS grows large. Regarding these phenomena, one reason is that the UatF bound relies on channel hardening and less hardening occurs when the spatial channel correlation is strong. The other reason is that channel estimate might be inaccurate at the strong spatial correlation and quantization errors, as shown in Fig. \ref{fig:NMSEvsASDAwareUnaware}.

\begin{figure}[!t] %[htb]
\centering
  \includegraphics[width=2.7in]{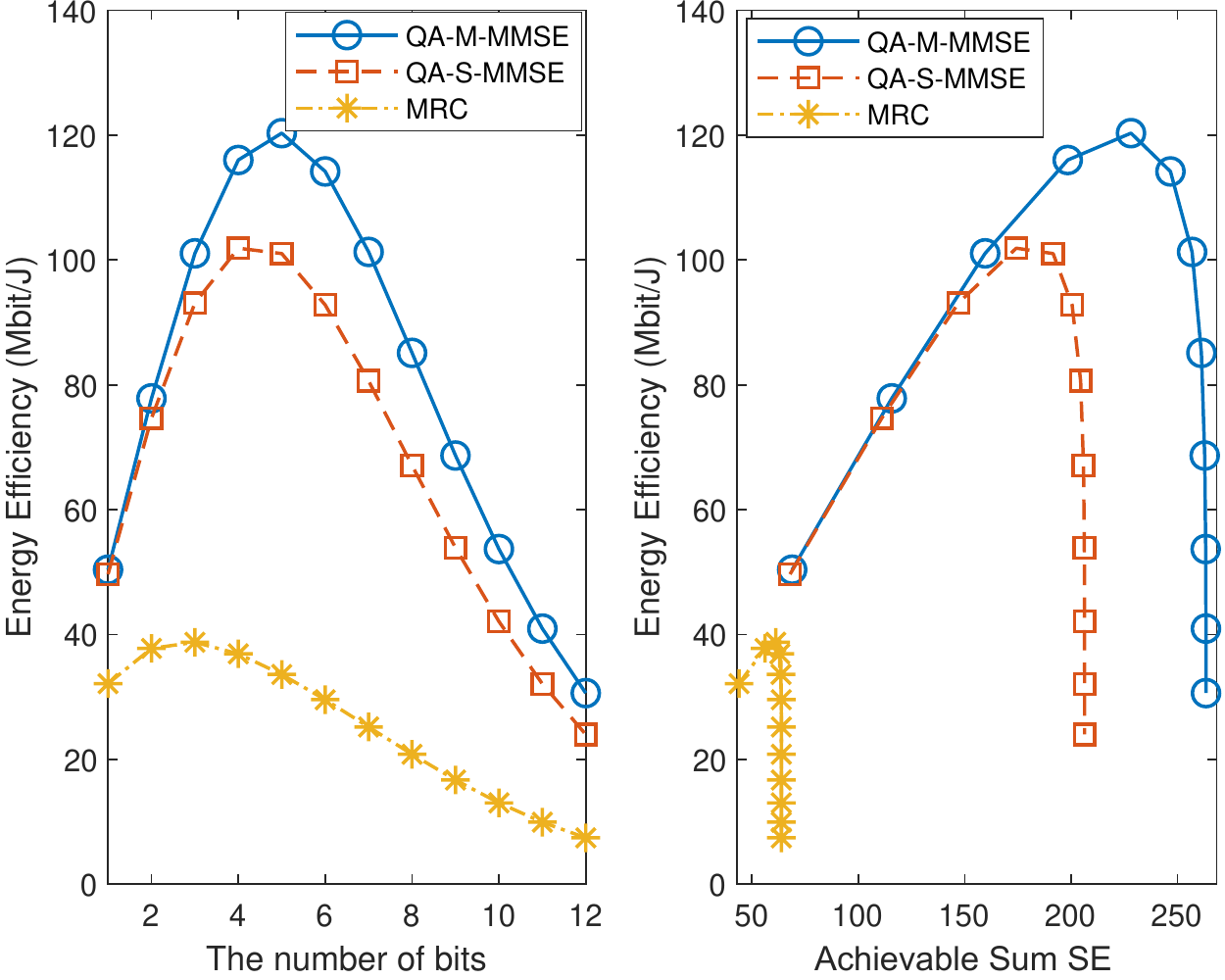}
%  \vspace{2.0cm}
%  \medskip
  \caption{Energy efficiency versus quantization bits (left) and the trade-off between achievable SE and energy efficiency (right). $L=9$, $M=30$, $K=5$, $f=3$, $\sigma_\varphi=10^\circ$, and $p=30$ dBm. \label{fig:EEvsBitRateMMSEMRC}}
\end{figure}

Fig. \ref{fig:EEvsBitRateMMSEMRC} presents the energy efficiency in conjunction with the trade-off between achievable sum SE and energy efficiency of multicell massive MIMO systems. Based on the existing work \cite{JLZBO2019}, the power consumption of different hardware components in (\ref{eq:totalPower}) are given as $P_{\mathrm{mix}}=30.3$ mW, $P_{\mathrm{filt}}=P_{\mathrm{filr}}=2.5$ mW, $P_{\mathrm{syn}}=50$ mW, $P_{\mathrm{LNA}}=20$ mW, $P_{\mathrm{IFA}}=3$ mW, $P_{\mathrm{AGC}}=2$ mW. As the number of quantization bits increases, it can be seen from the left part of Fig. \ref{fig:EEvsBitRateMMSEMRC} that the energy efficiencies of different combiners rise first but then decrease. This shows that the energy efficiencies of the three combiners have their own peaks, This suggests that it is not always advisable to increase the ADC quantization bits. Moreover, it can be observed from the right part of Fig. \ref{fig:EEvsBitRateMMSEMRC} that after a specific point slight increase of achievable sum SEs can result in lager decrease of energy efficiency, especially in the case where the achievable sum SEs are achieved under high-resolution ADCs. Based on these observations, it can be concluded that a multicell massive MIMO system with low-resolution ADCs (e.g., 3-5 bits) is more advantageous from the perspective of energy efficiency.

\section{Conclusion}
\label{sec:conclusion}

In this study, focusing on multicell massive MIMO systems with variable-resolution ADCs and spatially correlated Rayleigh fading channels, we derived the asymptotic closed-form expressions of the achievable uplink SE given that MRC and quantization-aware MMSE combiners are used at the BS. The tightness of our asymptotic analyses, which comprehensively consider the intra-cell and inter-cell interference, estimation errors, quantization noise, and  spatial correlation, is validated by simulation results. Among these results, we find that the proposed quantization-aware estimator and combiners are more helpful than the quantization-unaware counterparts. In addition, it is also verified that the performance of QA-M-MMSE outperforms that of QA-S-MMSE via considering the influence of inter-cell interference. Furthermore, we conclude that applying low-resolution ADCs in a multicell massive MIMO system is a promising candidate to achieve high energy efficiency. As future work, we intend to solve optimization problems under the constraints of power consumption and total quantization bits by regarding the asymptotic achievable SEs as objective functions. This is challenging because the quantization bits are discrete integers and the objective functions are non-convex.

% or


\appendices

\section{Basic Lemmas}
\label{app:someLemmas}
This appendix provides some basic lemmas that are helpful to derive all the principal SE bounds in this article.
\begin{Lemma}[Matrix Inversion Lemma] \label{lemma:MP}
Suppose that $\mathbf{A}$ is a  Hermitian invertible matrix and $\mathbf{A}+\alpha\mathbf{b}\mathbf{b}^H$ is also invertible for any vector $\mathbf{b}$ and  scalar $\alpha$. It holds that
\begin{equation}
  \mathbf{b}^H\left(\mathbf{A}+\alpha\mathbf{b}\mathbf{b}^H\right)^{-1}=\frac{\mathbf{b}^H\mathbf{A}^{-1}}{1+\alpha\mathbf{b}^H\mathbf{A}^{-1}\mathbf{b}}.
\end{equation}
\end{Lemma}

\begin{Lemma}\label{lemma:xAx}
Suppose that $\mathbf{A}$ has uniformly bounded spectral norm and $\mathbf{x}\thicksim \mathcal{CN}(\mathbf{0},\frac{1}{M}\mathbf{R}_\mathbf{x})$. It follows form \cite[Theorem 3.4]{Couillet:book} that $\mathbf{x}^H\mathbf{A}\mathbf{x}\xrightarrow[M\rightarrow\infty]{a.s.}\frac{1}{M}\mathrm{tr}(\mathbf{R}_\mathbf{x}\mathbf{A})$.
\end{Lemma}

\begin{Lemma} \label{lemma:AB}
  Suppose that $\mathbf{H}_i=[\mathbf{h}_{i1},\cdots,\mathbf{h}_{iK}]\in\mathbb{C}^{M\times K}$ and $\mathbf{H}=[\mathbf{H}_{1},\cdots,\mathbf{H}_{L}]\in\mathbb{C}^{M\times LK}$ with $\mathbf{h}_{ik^\prime}\thicksim \mathcal{CN}(\mathbf{0},\frac{1}{M}\mathbf{\Delta}_{ik^\prime})$ for $i=1,\cdots,L$ and $k^\prime=1,\cdots,K$. Suppose that $\mathbf{A}\in\mathbb{C}^{M\times M}$ and $\mathbf{D}\in\mathbb{C}^{M\times M}$ are nonnegative definite Hermitian matrices. Moreover, $\mathbf{A}$, $\mathbf{D}$, and $\mathbf{\Delta}_{ik^\prime}$ have bounded spectral norms. Then, for any positive $\alpha$, it follows from \cite[Theorem 1]{SRMD2012} that
  \begin{equation}
    \frac{1}{M}\mathrm{tr}\left(\mathbf{A}\left(\mathbf{H}\mathbf{H}^H+\mathbf{D}+\alpha\mathbf{I}_M\right)^{-1}\right)
    \xrightarrow[M\rightarrow\infty]{a.s.}\frac{1}{M}\mathrm{tr}\left(\mathbf{A}\bm{\Gamma}\right),
  \end{equation}
where $\bm{\Gamma}$ is given by
\begin{equation}\label{eq:Gamma}
  \bm{\Gamma}=\left(\frac{1}{M}\sum_{i=1}^{L}\sum_{k^\prime=1}^{K}\frac{\mathbf{\Delta}_{ik^\prime}}{1+\delta_{ik^\prime}}+
  \mathbf{D}+\alpha\mathbf{I}_M\right)^{-1}
\end{equation}
with $\delta_{ik^\prime}$ being the solution of fixed-point equation $\delta_{ik^\prime}=\frac{1}{M}\mathrm{tr}\left(\mathbf{\Delta}_{ik^\prime}\bm{\Gamma}\right)$.
\end{Lemma}

\begin{Lemma} \label{lemma:ABC}
  On the basis of Lemma \ref{lemma:AB}, for a nonnegative definite Hermitian matrix $\mathbf{C}\in\mathbb{C}^{M\times M}$ with bounded spectral norm, it follows from \cite{SRMD2012} that
  \begin{equation}
    \begin{split}
        & \frac{1}{M}\mathrm{tr}\left(\mathbf{A}\left(\mathbf{H}\mathbf{H}^H+\mathbf{D}+\alpha\mathbf{I}_M\right)^{-1}
  \mathbf{C}\left(\mathbf{H}\mathbf{H}^H+\mathbf{D}+\alpha\mathbf{I}_M\right)^{-1}\right) \\
         & \xrightarrow[M\rightarrow\infty]{a.s.}\frac{1}{M}\mathrm{tr}\left(\mathbf{A}\bm{\Gamma}^\prime\right).
    \end{split}
  \end{equation}
Moreover, $\bm{\Gamma}^\prime$ is given by
\begin{equation}\label{eq:GammaPrime}
  \bm{\Gamma}^\prime=\bm{\Gamma}\mathbf{C}\bm{\Gamma}+\frac{1}{M}\bm{\Gamma}\sum_{i=1}^{L}\sum_{k^\prime=1}^{K}
  \frac{\mathbf{\Delta}_{ik^\prime}\delta_{ik^\prime}^\prime}{(1+\delta_{ik^\prime})^2}\bm{\Gamma}.
\end{equation}
If letting $\bm{\delta}^\prime=[(\bm{\delta}^\prime_1)^T,\cdots,(\bm{\delta}^\prime_L)^T]^T$ with $\bm{\delta}^\prime_i=[\delta^\prime_{i1},\cdots,\delta^\prime_{iK}]^T$, we have $\bm{\delta}^\prime=(\mathbf{I}_{LK}-\mathbf{Y})^{-1}\mathbf{x}$ with the elements of $\mathbf{Y}\in\mathbb{C}^{LK\times LK}$ and $\mathbf{x}\in\mathbb{C}^{LK\times 1}$ given by
\begin{equation}
  [\mathbf{Y}]_{ik^\prime,\bar{i}\bar{k}^\prime}=\frac{\frac{1}{M}\mathrm{tr}(\mathbf{\Delta}_{ik^\prime}\bm{\Gamma}\mathbf{\Delta}_{\bar{i}\bar{k}^\prime}\bm{\Gamma})}{M(1+\delta_{\bar{i}\bar{k}^\prime})^2}
\end{equation}
and
\begin{equation}
  [\mathbf{x}]_{ik^\prime}=\frac{1}{M}\mathrm{tr}(\mathbf{\Delta}_{ik^\prime}\bm{\Gamma}\mathbf{C}\bm{\Gamma}).
\end{equation}
\end{Lemma}

\begin{Lemma}[Rank-1 Perturbation Lemma] \label{lemma:perturbationlemma}
Suppose that $\mathbf{A}\in\mathbb{C}^{M\times M}$, $\mathbf{q}\in\mathbb{C}^{M\times 1}$, and that $\mathbf{G}\in\mathbb{C}^{M\times M}$ is a nonnegative Hermitian matrix. With $\alpha$ and $\alpha^{\prime}$ two given positive real numbers, it follows form \cite[Theorem 3.9]{Couillet:book} that
\begin{equation}
  \left|\mathrm{tr}\left(\mathbf{A}(\mathbf{G}+\alpha\mathbf{I}_M)^{-1}-\mathbf{A}(\mathbf{G}+\alpha^{\prime}\mathbf{g}\mathbf{g}^H+\alpha\mathbf{I}_M)^{-1}\right)\right|
\leq \frac{||\mathbf{A}||}{\alpha}.
\end{equation}
\end{Lemma}

\begin{Lemma} \label{lemma:aBa}
Suppose that $\mathbf{a}\thicksim \mathcal{CN}(\mathbf{0},\mathbf{A})$ and that $\mathbf{B}\in\mathbb{C}^{M\times M}$ is a diagonalizable matrix. It holds that
\begin{equation}
  \mathbb{E}\{|\mathbf{a}^H\mathbf{B}\mathbf{a}|^2\}=|\tr(\mathbf{B}\mathbf{A})|^2+\tr(\mathbf{B}\mathbf{A}\mathbf{B}^H\mathbf{A}).
\end{equation}
\end{Lemma}

\section{Proof of Theorem \ref{lemma:SEMRC}}
\label{app:lemmaMRC}

It is straightforward to obtain $A_{j,k}$ and $F_{j,k}$ following Lemma \ref{lemma:xAx}. In this appendix, we will derive the closed-form expressions of $\Psi_{j,k}=B_{j,k}+C_{j,k}+D_{j,k}+E_{j,k}$ and $G_{j,k}$.

1) We first deal with $\Psi_{j,k}$. According to (\ref{eq:ExpeBjk})-(\ref{eq:ExpeEjk}) and ${\mathbf{h}}_{j,ik^\prime}=\hat{\mathbf{h}}_{j,ik^\prime}+\tilde{\mathbf{h}}_{j,ik^\prime}$, it follows that
\begin{equation}\label{eq:Psi0}
  \Psi_{j,k}=\sum_{i=1}^{L}\sum_{k^\prime=1}^{K}p_{i,k^\prime}\mathbb{E}\{|\mathbf{v}_{j,k}^H\mathbf{h}_{j,ik^\prime}|^2\}-A_{j,k}.
\end{equation}
If we let $\psi_{i,k^\prime}=\mathbb{E}\{|\mathbf{v}_{j,k}^H\mathbf{h}_{j,ik^\prime}|^2\}$, the following two cases are discussed because of the pilot contamination.

i) When $(i,k^\prime)\notin \mathcal{P}_{j,k}$, $\hat{\mathbf{h}}_{j,jk}$ and $\mathbf{h}_{j,ik^\prime}$ are uncorrelated with each other. In this case, it follows that
\begin{equation}\label{eq:psinotin}
\begin{split}
   \psi_{i,k^\prime} & =\mathbb{E}\{|\hat{\mathbf{h}}_{j,jk}^H\mathbf{h}_{j,ik^\prime}|^2\}\\
   &=\mathbb{E}\{\hat{\mathbf{h}}_{j,jk}^H\mathbb{E}\{\mathbf{h}_{j,ik^\prime}\mathbf{h}_{j,ik^\prime}^H\}\hat{\mathbf{h}}_{j,jk}\}\\
     & =\tr(\mathbb{E}\{\mathbf{h}_{j,ik^\prime}\mathbf{h}_{j,ik^\prime}^H\}\mathbb{E}\{\hat{\mathbf{h}}_{j,jk}\hat{\mathbf{h}}_{j,jk}^H\})\\
     &=\tr(\mathbf{R}_{j,ik^\prime}\mathbf{B}_{j,jk}).
\end{split}
\end{equation}

ii) When $(i,k^\prime)\in \mathcal{P}_{j,k}$, $\hat{\mathbf{h}}_{j,jk}$ and $\hat{\mathbf{h}}_{j,ik^\prime}$ are uncorrelated with each other. In this regard, $\psi_{i,k^\prime}$ can be decomposed as
\begin{equation}\label{eq:psi0}
  \psi_{i,k^\prime}=\mathbb{E}\{|\hat{\mathbf{h}}_{j,jk}^H\hat{\mathbf{h}}_{j,ik^\prime}|^2\}+\mathbb{E}\{|\hat{\mathbf{h}}_{j,jk}^H\tilde{\mathbf{h}}_{j,ik^\prime}|^2\}.
\end{equation}
We first focus on the first term of the RHS in (\ref{eq:psi0}) and let $\varphi_{i,k^\prime} =\mathbb{E}\{|\hat{\mathbf{h}}_{j,jk}^H\hat{\mathbf{h}}_{j,ik^\prime}|^2\}$. According to (\ref{eq:CSIcorr}), one has
\begin{equation}\label{eq:psihhat}
   \varphi_{i,k^\prime} =\frac{{p_{i,k^\prime}}}{{p_{j,k}}}\mathbb{E}\left\{\left|\hat{\mathbf{h}}_{j,jk}^H\mathbf{R}_{j,i k^\prime}(\mathbf{R}_{j,jk})^{-1}\hat{\mathbf{h}}_{j,j k}\right|^2\right\}.
\end{equation}
Based on Lemma \ref{lemma:aBa}, it holds that
\begin{equation}\label{eq:psihhat2}
  \begin{split}
   \varphi_{i,k^\prime} & =\frac{{p_{i,k^\prime}}}{{p_{j,k}}}\bigg[\left|\tr(\mathbf{R}_{j,i k^\prime}(\mathbf{R}_{j,jk})^{-1}\mathbf{B}_{j,jk})\right|^2 \\
     & +\tr(\mathbf{R}_{j,i k^\prime}(\mathbf{R}_{j,jk})^{-1}\mathbf{B}_{j,jk}(\mathbf{R}_{j,jk})^{-1}\mathbf{R}_{j,i k^\prime}\mathbf{B}_{j,jk})\bigg].
\end{split}
\end{equation}
Using $\mathbf{B}_{j,ik^\prime}=p_{i,k^\prime}\tau_p\mathbf{R}_{j,ik^\prime}\bm{\Sigma}^\AD_j\bm{\Psi}_{j,ik^\prime}\bm{\Sigma}^\AD_j\mathbf{R}_{j,ik^\prime}$ given by (\ref{eq:covarianceBjik}), $\varphi_{i,k^\prime}$ can be rewritten as
\begin{equation}\label{eq:psihhat3}
  \begin{split}
   \varphi_{i,k^\prime} & =p_{i k^\prime}p_{j k}\tau_p^2\left|\tr(\mathbf{R}_{j,i k^\prime}\bm{\Sigma}^\AD_j\bm{\Psi}_{j,jk}\bm{\Sigma}^\AD_j\mathbf{R}_{j,jk})\right|^2 \\
     & +\tr(\mathbf{B}_{j,ik^\prime}\mathbf{B}_{j,jk}).
\end{split}
\end{equation}
Notice that $\bm{\Psi}_{j,jk}=\bm{\Psi}_{j,ik^\prime}$ in the case of $(i,k^\prime)\in \mathcal{P}_{j,k}$.
Next, similarly to the derivation of (\ref{eq:psinotin}), it holds for the second term of the RHS in (\ref{eq:psi0}) that
\begin{equation}\label{eq:psihtilde}
  \mathbb{E}\{|\hat{\mathbf{h}}_{j,jk}^H\tilde{\mathbf{h}}_{j,ik^\prime}|^2\}=\tr(\mathbf{B}_{j,jk}\mathbf{C}_{j,ik^\prime})
\end{equation}
since $\hat{\mathbf{h}}_{j,ik^\prime}$ and $\tilde{\mathbf{h}}_{j,ik^\prime}$ are mutually uncorrelated. Substituting (\ref{eq:psihtilde}) and (\ref{eq:psihhat3}) into (\ref{eq:psi0}) with the aid of $\mathbf{R}_{j,ik^\prime}=\mathbf{B}_{j,ik}+\mathbf{C}_{j,ik^\prime}$ leads to
\begin{equation}\label{eq:psihhat4}
  \begin{split}
   \psi_{i,k^\prime} & =p_{i,k^\prime}p_{j,k}\tau_p^2\left|\tr(\mathbf{R}_{j,i k^\prime}\bm{\Sigma}^\AD_j\bm{\Psi}_{j,jk}\bm{\Sigma}^\AD_j\mathbf{R}_{j,jk})\right|^2 \\
     & +\tr(\mathbf{R}_{j,ik^\prime}\mathbf{B}_{j,jk}).
\end{split}
\end{equation}
Combining the results of (\ref{eq:psinotin}) and (\ref{eq:psihhat4}), we can obtain (\ref{eq:BjkEjikMRC}).

2) Subsequently, we try to derive the closed-form expression of $G_{j,k}$. Based on (\ref{eq:quantizationCovarianceData}) and (\ref{eq:ExpeGjk}), we have
\begin{equation}\label{eq:ExpeGjk1}
\begin{split}
   G_{j,k} & =\sum_{i=1}^{L}\sum_{k^\prime=1}^{K}\sum_{m=1}^{M}\frac{1-\alpha^\AD_{j,m}}{\alpha^\AD_{j,m}}
    \kappa^m_{j,ik^\prime}+\mu_{j,k},
\end{split}
\end{equation}
where $\kappa^m_{j,ik^\prime}=\mathbb{E}\{p_{i,k^\prime}|{h}^m_{j,ik^\prime}|^2|\hat{h}^m_{j,jk}|^2\}$ and $\mu_{j,k}=\sigma^2\mathrm{tr}(\mathbf{B}_{j,jk}((\bm{\Sigma}^\AD_j)^{-1}-\mathbf{I}_M))$. Owing to the pilot contamination, two cases are discussed as follows.

i) When $(i,k^\prime)\notin \mathcal{P}_{j,k}$, ${h}^m_{j,ik^\prime}$ and $\hat{h}^m_{j,jk}$ are uncorrelated with each other. In this case, it follows that
\begin{equation}\label{eq:Kappa1}
  \kappa^m_{j,ik^\prime}=p_{i,k^\prime}(\mathbf{R}_{j,ik^\prime})_{mm}(\mathbf{B}_{j,jk})_{mm}.
\end{equation}
Inserting (\ref{eq:Kappa1}) into (\ref{eq:ExpeGjk1}) yields $G_{j,k}$ for this case.

ii) When $(i,k^\prime)\in \mathcal{P}_{j,k}$, $\hat{h}^m_{j,ik^\prime}$ and $\hat{h}^m_{j,jk}$ are correlated with each other. By Letting $\chi^m_{j,ik^\prime}=\mathbb{E}\{|\hat{h}^m_{j,ik^\prime}|^2|\hat{h}^m_{j,jk}|^2\}$, $\kappa^m_{j,ik^\prime}$ is further decomposed as
\begin{equation}\label{eq:kappa0}
  \begin{split}
     \kappa^m_{j,ik^\prime} & =p_{i,k^\prime}\chi^m_{j,ik^\prime} + p_{i,k^\prime}\mathbb{E}\{|\tilde{h}^m_{j,ik^\prime}|^2|\hat{h}^m_{j,jk}|^2\}  \\
       & =p_{i,k^\prime}\chi^m_{j,ik^\prime} +p_{i,k^\prime}(\mathbf{C}_{j,ik^\prime})_{mm}(\mathbf{B}_{j,jk})_{mm}
  \end{split}
\end{equation}
Recalling from (\ref{eq:corrh1h2}) in the case of $(i,k^\prime)\in \mathcal{P}_{j,k}$, we have
\begin{equation*}
  \hat{\mathbf{h}}_{j,i k^\prime}=\frac{\sqrt{p_{i,k^\prime}}}{\sqrt{p_{j,k}}}\mathbf{R}_{j,i k^\prime}(\mathbf{R}_{j,jk})^{-1}\hat{\mathbf{h}}_{j,j k}
\end{equation*}
with which we can obtain
\begin{equation}\label{eq:hath2}
  \begin{split}
     &|\hat{h}^m_{j,ik^\prime}|^2  =(\hat{\mathbf{h}}_{j,i k^\prime}\hat{\mathbf{h}}^H_{j,i k^\prime})_{mm} \\
       & =\frac{{p_{i,k^\prime}}}{{p_{j,k}}}\left(\mathbf{R}_{j,i k^\prime}(\mathbf{R}_{j,jk})^{-1}\hat{\mathbf{h}}_{j,j k}\hat{\mathbf{h}}_{j,j k}^H(\mathbf{R}_{j,jk})^{-1}\mathbf{R}_{j,i k^\prime}\right)_{mm}.
  \end{split}
\end{equation}
Based on (\ref{eq:hath2}) and $\chi^m_{j,ik^\prime}=\mathbb{E}\{|\hat{h}^m_{j,ik^\prime}|^2|\hat{h}^m_{j,jk}|^2\}$, we have
\begin{equation}\label{eq:tmp0}
  \chi^m_{j,ik^\prime}=\frac{{p_{i,k^\prime}}}{{p_{j,k}}}\left(\mathbf{R}_{j,i k^\prime}(\mathbf{R}_{j,jk})^{-1}\mathbf{V}^m_{j,jk}(\mathbf{R}_{j,jk})^{-1}\mathbf{R}_{j,i k^\prime}\right)_{mm},
\end{equation}
where the $(m_1,m_2)$th component of $\mathbf{V}^m_{j,jk}$ is given by
\begin{equation*}
  (\mathbf{V}^m_{j,jk})_{m_1m_2}=\mathbb{E}\{\hat{h}_{j,jk}^{m_1}(\hat{h}_{j,jk}^{m_2})^\ast\hat{h}_{j,jk}^{m}(\hat{h}_{j,jk}^{m})^\ast\}.
\end{equation*}
If letting $\mathbf{T}=\mathbf{B}_{j,jk}^{1/2}$ and using the channel model in (\ref{eq:correlatedCSI}), one has the $m$th element of $\hat{h}_{j,jk}$ according to
\begin{equation}\label{eq:hscalar}
  h^m_{j,jk}=\sum_{\bar{m}=1}^{M}t_{m\bar{m}}x^{\bar{m}}_{j,jk},
\end{equation}
with which we can arrive at the result of (\ref{eq:Vmjjk}). Inserting (\ref{eq:tmp0}) back into (\ref{eq:kappa0}) first and then plugging (\ref{eq:kappa0}) back into (\ref{eq:ExpeGjk1}) lead to $G_{j,k}$ for the second case. Combining the above two cases, we can obtain (\ref{eq:GjkMRC}). With all terms having been derived, we complete the proof of Theorem \ref{lemma:SEMRC}.

\section{Proof of Theorem \ref{lemma:SEMMMSE}}
\label{app:lemmaMMMSE}

For brevity during the following derivations, we first define
\begin{equation}\label{eq:Lambdaj}
  \bm{\Lambda}_j=\frac{1}{Mp_{j,k}}\left(\sum_{i=1}^{L}\hat{\mathbf{H}}_{j,i}\mathbf{P}_{i}(\hat{\mathbf{H}}_{j,i})^H+\mathbf{Z}_j^\mathrm{M}+\sigma^2\mathbf{I}_M\right),
\end{equation}
\begin{equation}\label{eq:Lambdajik}
  \bm{\Lambda}_{j,ik^\prime}=\bm{\Lambda}_j-\frac{p_{i,k^\prime}}{Mp_{jk}}\hat{\mathbf{h}}_{j,ik^\prime}\hat{\mathbf{h}}_{j,ik^\prime}^H,
\end{equation}
and
\begin{equation}\label{eq:Lambdajiklm}
  \bm{\Lambda}_{j,ik^\prime,l\bar{k}}=\bm{\Lambda}_{j,ik^\prime}-\frac{p_{l,\bar{k}}}{Mp_{jk}}\hat{\mathbf{h}}_{j,l\bar{k}}\hat{\mathbf{h}}_{j,l\bar{k}}^H
\end{equation}
for $(l,\bar{k})\neq(i,k^\prime)$. Based on the definition of $\bm{\Lambda}_j$, $\mathbf{v}_{j,k}^{\mathrm{M}}$ in (\ref{eq:vjkMMMSE}) can be rewritten as $\mathbf{v}_{j,k}^{\mathrm{M}}=\frac{1}{M}\bm{\Lambda}_j^{-1}\hat{\mathbf{h}}_{j,jk}$.

1) Compute asymptotic $A_{j,k}$: If we let $a_k=\mathbb{E}\{(\mathbf{v}_{j,k}^{\mathrm{M}})^H\hat{\mathbf{h}}_{j,jk}\}$, it follows that
\begin{equation}\label{eq:ajk}
  \begin{split}
     a_k&= \mathbb{E}\left\{\frac{\hat{\mathbf{h}}_{j,jk}^H\bm{\Lambda}_j^{-1}\hat{\mathbf{h}}_{j,jk}}{M}\right\}
\overset{(a)}{=} \mathbb{E}\left\{\frac{\frac{\hat{\mathbf{h}}_{j,jk}^H\bm{\Lambda}_{j,jk}^{-1}\hat{\mathbf{h}}_{j,jk}}{M}}
{1+\frac{\hat{\mathbf{h}}_{j,jk}^H\bm{\Lambda}_{j,jk}^{-1}\hat{\mathbf{h}}_{j,jk}}{M}}\right\}\\
       &\overset{(b)}{\approx} \mathbb{E}\left\{\frac{\frac{\hat{\mathbf{h}}_{j,jk}^H\bm{\Lambda}_{j}^{-1}\hat{\mathbf{h}}_{j,jk}}{M}}
{1+\frac{\hat{\mathbf{h}}_{j,jk}^H\bm{\Lambda}_{j}^{-1}\hat{\mathbf{h}}_{j,jk}}{M}}\right\}
\overset{(c)}{\approx} \frac{\frac{\tr(\mathbf{B}_{j,jk}\bm{\Lambda}^{-1}_j)}{M}}
{1+\frac{\tr(\mathbf{B}_{j,jk}\bm{\Lambda}^{-1}_j)}{M}}\\
& \overset{(d)}{\approx}\frac{\frac{\mathrm{tr}(\mathbf{B}_{j,jk}\mathbf{T}_{j,jk})}{M}}{1+\frac{\mathrm{tr}(\mathbf{B}_{j,jk}\mathbf{T}_{j,jk})}{M}}.
  \end{split}
\end{equation}
Note that $(a)$ follows Lemma \ref{lemma:MP}, $(b)$ utilizes Lemma \ref{lemma:perturbationlemma}, $(c)$ is obtained via Lemma \ref{lemma:xAx}, and $(d)$ is achieved via Lemma \ref{lemma:AB} by letting $\mathbf{A}=\mathbf{B}_{j,jk}$, $\bm{\Delta}_{ik^\prime}=\frac{p_{i,k^\prime}}{p_{j,k}}\mathbf{B}_{j,ik^\prime}$, $\mathbf{D}=\frac{\mathbf{Z}^{\mathrm{M}}_{j}}{p_{j,k}M}$, and $\alpha=\frac{\sigma^2}{p_{j,k}M}$. Due to $A_{j,k}= p_{j,k}|a_k|^2$, we can come to (\ref{eq:AjkMMMSE}) following (\ref{eq:ajk}). Based on the derivations of $A_{j,k}$, it is easy to obtain that $B_{j,k}\rightarrow p_{j,k}|a_k|^2-A_{j,k}=0$.

2) Compute asymptotic $B_{j,k}$: If we let $c_{k^\prime}=\mathbb{E}\{|(\mathbf{v}_{j,k}^{\mathrm{M}})^H\hat{\mathbf{h}}_{j,jk^\prime}|^2\}$, $c_{k^\prime}$ can be deduced as
\begin{equation}\label{eq:Bjk}
  \begin{split}
     &c_{k^\prime}= \frac{1}{M^2} \mathbb{E}\left\{\hat{\mathbf{h}}_{j,jk}^H\bm{\Lambda}_j^{-1}\hat{\mathbf{h}}_{j,jk^\prime}\hat{\mathbf{h}}_{j,jk^\prime}^H\bm{\Lambda}_j^{-1}\hat{\mathbf{h}}_{j,jk}\right\}\\
&\overset{(a)}{=} \frac{1}{M^2}\mathbb{E}\left\{\frac{\hat{\mathbf{h}}_{j,jk}^H\bm{\Lambda}_{j,jk}^{-1}\hat{\mathbf{h}}_{j,jk^\prime}\hat{\mathbf{h}}_{j,jk^\prime}^H\bm{\Lambda}_{j,jk}^{-1}\hat{\mathbf{h}}_{j,jk}}
{|1+\frac{1}{M}\hat{\mathbf{h}}_{j,jk}^H\bm{\Lambda}_{j,jk}^{-1}\hat{\mathbf{h}}_{j,jk}|^2}\right\}\\
       &\overset{(b)}{=} \frac{1}{M^2}\mathbb{E}\left\{\frac{\hat{\mathbf{h}}_{j,jk}^H\bm{\Lambda}_{j,jk}^{-1}\hat{\mathbf{h}}_{j,jk^\prime}\hat{\mathbf{h}}_{j,jk^\prime}^H\bm{\Lambda}_{j,jk}^{-1}\hat{\mathbf{h}}_{j,jk}}
{|1+\frac{\hat{\mathbf{h}}_{j,jk}^H\bm{\Lambda}_{j,jk}^{-1}\hat{\mathbf{h}}_{j,jk}}{M}|^2|1+\frac{\hat{\mathbf{h}}_{j,jk^\prime}^H\bm{\Lambda}_{j,jk,jk^\prime}^{-1}\hat{\mathbf{h}}_{j,jk^\prime}}{M}|^2}\right\}\\
&\overset{(c)}{\approx} \frac{1}{M^2}\mathbb{E}\left\{\frac{\hat{\mathbf{h}}_{j,jk}^H\bm{\Lambda}_{j}^{-1}\hat{\mathbf{h}}_{j,jk^\prime}\hat{\mathbf{h}}_{j,jk^\prime}^H\bm{\Lambda}_{j}^{-1}\hat{\mathbf{h}}_{j,jk}}
{|1+\frac{\hat{\mathbf{h}}_{j,jk}^H\bm{\Lambda}_{j}^{-1}\hat{\mathbf{h}}_{j,jk}}{M}|^2|1+\frac{\hat{\mathbf{h}}_{j,jk^\prime}^H\bm{\Lambda}_{j}^{-1}\hat{\mathbf{h}}_{j,jk^\prime}}{M}|^2}\right\}\\
& \overset{(d)}{\approx}\frac{1}{M^2}\frac{\tr(\mathbf{B}_{j,jk}\bm{\Lambda}_{j}^{-1}\mathbf{B}_{j,jk^\prime}\bm{\Lambda}_{j}^{-1})}
{|1+\frac{\tr(\mathbf{B}_{j,jk}\bm{\Lambda}_{j}^{-1})}{M}|^2|1+\frac{\tr(\mathbf{B}_{j,jk^\prime}\bm{\Lambda}_{j}^{-1})}{M}|^2}\\
&\overset{(e)}{\approx}\frac{\frac{\mathrm{tr}(\mathbf{B}_{j,jk}\mathbf{T}^{\prime}_{j,jk^\prime})}{M^2}}{\left|1+\frac{\mathrm{tr}(\mathbf{B}_{j,jk}\mathbf{T}_{j,jk})}{M}\right|^2\left|1+\frac{\mathrm{tr}(\mathbf{B}_{j,jk^\prime}\mathbf{T}_{j,jk^\prime})}{M}\right|^2}.
  \end{split}
\end{equation}
Note that $(a)$ and $(b)$ exploit Lemma \ref{lemma:MP}, $(c)$ utilizes Lemma \ref{lemma:perturbationlemma}, $(d)$ is obtained via Lemma \ref{lemma:xAx}, and $(e)$ is achieved via Lemma \ref{lemma:ABC} by letting $\mathbf{C}=\mathbf{B}_{j,jk^\prime}$. Due to $C_{j,k}=\sum_{k\neq k^\prime}^{K} p_{j,k^\prime}c_{k^\prime}$, we can obtain (\ref{eq:CjkMMMSE}) based on (\ref{eq:Bjk}).

3) Compute asymptotic $D_{j,k}$: The derivations of $D_{j,k}$ are similar to that of $C_{j,k}$. The difference is that we let $\mathbf{C}=\mathbf{B}_{j,ik^\prime}$ in Lemma \ref{lemma:ABC}.

4) Compute asymptotic $E_{j,k}$: If we let $e_{ik^\prime}=\mathbb{E}\{|(\mathbf{v}_{j,k}^{\mathrm{M}})^H\tilde{\mathbf{h}}_{j,ik^\prime}|^2\}$, it follows that
\begin{equation}\label{eq:Ejk}
  \begin{split}
     e_{ik^\prime}&= \frac{1}{M^2} \mathbb{E}\left\{\hat{\mathbf{h}}_{j,jk}^H\bm{\Lambda}_j^{-1}\tilde{\mathbf{h}}_{j,ik^\prime}\tilde{\mathbf{h}}_{j,ik^\prime}^H\bm{\Lambda}_j^{-1}\hat{\mathbf{h}}_{j,jk}\right\}\\
&\overset{(a)}{=} \frac{1}{M^2}\mathbb{E}\left\{\frac{\hat{\mathbf{h}}_{j,jk}^H\bm{\Lambda}_{j,jk}^{-1}\tilde{\mathbf{h}}_{j,ik^\prime}\tilde{\mathbf{h}}_{j,ik^\prime}^H\bm{\Lambda}_{j,jk}^{-1}\hat{\mathbf{h}}_{j,jk}}
{|1+\frac{1}{M}\hat{\mathbf{h}}_{j,jk}^H\bm{\Lambda}_{j,jk}^{-1}\hat{\mathbf{h}}_{j,jk}|^2}\right\}\\
&\overset{(b)}{\approx}\frac{1}{M^2}\mathbb{E}\left\{\frac{\hat{\mathbf{h}}_{j,jk}^H\bm{\Lambda}_{j}^{-1}\tilde{\mathbf{h}}_{j,ik^\prime}\tilde{\mathbf{h}}_{j,ik^\prime}^H\bm{\Lambda}_{j}^{-1}\hat{\mathbf{h}}_{j,jk}}
{|1+\frac{1}{M}\hat{\mathbf{h}}_{j,jk}^H\bm{\Lambda}_{j}^{-1}\hat{\mathbf{h}}_{j,jk}|^2}\right\}\\
&\overset{(c)}{\approx}\frac{1}{M^2}\mathbb{E}\left\{\frac{\tr(\mathbf{B}_{j,jk}\bm{\Lambda}_{j}^{-1}\mathbf{C}_{j,ik^\prime}\bm{\Lambda}_{j}^{-1})}
{|1+\frac{1}{M}\tr(\mathbf{B}_{j,jk}\bm{\Lambda}_{j}^{-1})|^2}\right\}\\
&\overset{(d)}{\approx}\frac{\frac{\mathrm{tr}(\mathbf{B}_{j,jk}\tilde{\mathbf{T}}^{\prime}_{j,ik^\prime})}{M^2}}{\left|1+\frac{\mathrm{tr}(\mathbf{B}_{j,jk}\mathbf{T}_{j,jk})}{M}\right|^2}.
\end{split}
\end{equation}
Note that $(a)$ exploits Lemma \ref{lemma:MP}, $(b)$ utilizes Lemma \ref{lemma:perturbationlemma}, $(c)$ is obtained via Lemma \ref{lemma:xAx}, and $(d)$ is achieved via Lemma \ref{lemma:AB} by letting $\mathbf{C}=\mathbf{C}_{j,ik^\prime}$. By using (\ref{eq:Ejk}), $E_{j,k}$ turns into (\ref{eq:EjkMMMSE}) due to $E_{j,k}=\sum_{i=1}^{K}\sum_{k^\prime=1}^{K} p_{i,k^\prime}e_{ik^\prime}$.

5) Compute asymptotic $E_{j,k}$: It follows from $E_{j,k}=\sigma^2\mathbb{E}\{||\mathbf{v}_{j,k}||^2\}$ that
\begin{equation}\label{eq:Fjk}
  \begin{split}
     E_{j,k}&= \frac{\sigma^2}{M^2} \mathbb{E}\left\{\hat{\mathbf{h}}_{j,jk}^H\bm{\Lambda}_j^{-1}\bm{\Lambda}_j^{-1}\hat{\mathbf{h}}_{j,jk}\right\}\\
&\overset{(a)}{=} \frac{\sigma^2}{M^2}\mathbb{E}\left\{\frac{\hat{\mathbf{h}}_{j,jk}^H\bm{\Lambda}_{j,jk}^{-1}\bm{\Lambda}_{j,jk}^{-1}\hat{\mathbf{h}}_{j,jk}}
{|1+\frac{1}{M}\hat{\mathbf{h}}_{j,jk}^H\bm{\Lambda}_{j,jk}^{-1}\hat{\mathbf{h}}_{j,jk}|^2}\right\}\\
&\overset{(b)}{\approx}\frac{\sigma^2}{M^2}\mathbb{E}\left\{\frac{\hat{\mathbf{h}}_{j,jk}^H\bm{\Lambda}_{j}^{-1}\bm{\Lambda}_{j}^{-1}\hat{\mathbf{h}}_{j,jk}}
{|1+\frac{1}{M}\hat{\mathbf{h}}_{j,jk}^H\bm{\Lambda}_{j}^{-1}\hat{\mathbf{h}}_{j,jk}|^2}\right\}\\
&\overset{(c)}{\approx}\frac{\sigma^2}{M^2}\mathbb{E}\left\{\frac{\tr(\mathbf{B}_{j,jk}\bm{\Lambda}_{j}^{-1}\mathbf{I}_M\bm{\Lambda}_{j}^{-1})}
{|1+\frac{1}{M}\tr(\mathbf{B}_{j,jk}\bm{\Lambda}_{j}^{-1})|^2}\right\}\\
&\overset{(d)}{\approx}\frac{\sigma^2\frac{\mathrm{tr}(\mathbf{B}_{j,jk}\mathbf{T}^{\prime}_{jk,\mathrm{n}})}{M^2}}{\left|1+\frac{\mathrm{tr}(\mathbf{B}_{j,jk}\mathbf{T}_{j,jk})}{M}\right|^2}.
\end{split}
\end{equation}
Note that $(a)$ follows Lemma \ref{lemma:MP}, $(b)$ utilizes Lemma \ref{lemma:perturbationlemma}, $(c)$ is computed by applying Lemma \ref{lemma:xAx}, and $(d)$ is achieved via Lemma \ref{lemma:AB} by letting $\mathbf{C}=\mathbf{I}_M$.

6) Compute asymptotic $G_{j,k}$: It follows from $G_{j,k}=\mathbb{E}\{\mathbf{v}_{j,k}^H\mathbf{R}_{\mathbf{q}_j}\mathbf{v}_{j,k}\}$ that
\begin{equation}\label{eq:Gjk}
  \begin{split}
     &G_{j,k}= \frac{1}{M^2} \mathbb{E}\left\{\hat{\mathbf{h}}_{j,jk}^H\bm{\Lambda}_j^{-1}(\bm{\Sigma}^\AD_j)^{-1}\mathbf{R}_{\mathbf{q}_j}(\bm{\Sigma}^\AD_j)^{-1}\bm{\Lambda}_j^{-1}\hat{\mathbf{h}}_{j,jk}\right\}\\
&\overset{(a)}{=} \frac{1}{M^2}\mathbb{E}\left\{\frac{\hat{\mathbf{h}}_{j,jk}^H\bm{\Lambda}_{j,jk}^{-1}(\bm{\Sigma}^\AD_j)^{-1}\mathbf{R}_{\mathbf{q}_j}(\bm{\Sigma}^\AD_j)^{-1}\bm{\Lambda}_{j,jk}^{-1}\hat{\mathbf{h}}_{j,jk}}
{|1+\frac{1}{M}\hat{\mathbf{h}}_{j,jk}^H\bm{\Lambda}_{j,jk}^{-1}\hat{\mathbf{h}}_{j,jk}|^2}\right\}\\
&\overset{(b)}{\approx}\frac{1}{M^2}\mathbb{E}\left\{\frac{\hat{\mathbf{h}}_{j,jk}^H\bm{\Lambda}_{j}^{-1}(\bm{\Sigma}^\AD_j)^{-1}\mathbf{R}_{\mathbf{q}_j}(\bm{\Sigma}^\AD_j)^{-1}\bm{\Lambda}_{j}^{-1}\hat{\mathbf{h}}_{j,jk}}
{|1+\frac{1}{M}\hat{\mathbf{h}}_{j,jk}^H\bm{\Lambda}_{j}^{-1}\hat{\mathbf{h}}_{j,jk}|^2}\right\}\\
&\overset{(c)}{\approx}\frac{1}{M^2}\mathbb{E}\left\{\frac{\tr(\mathbf{B}_{j,jk}\bm{\Lambda}_{j}^{-1}(\bm{\Sigma}^\AD_j)^{-1}\mathbf{R}_{\mathbf{q}_j}(\bm{\Sigma}^\AD_j)^{-1}\bm{\Lambda}_{j}^{-1})}
{|1+\frac{1}{M}\tr(\mathbf{B}_{j,jk}\bm{\Lambda}_{j}^{-1})|^2}\right\}\\
&\overset{(d)}{\approx}\frac{\frac{\mathrm{tr}(\mathbf{B}_{j,jk}\mathbf{T}^{\prime}_{jk,\mathrm{q}})}{M^2}}{\left|1+\frac{\mathrm{tr}(\mathbf{B}_{j,jk}\mathbf{T}_{j,jk})}{M}\right|^2}.
\end{split}
\end{equation}
Note that $(a)$ follows Lemma \ref{lemma:MP}, $(b)$ utilizes Lemma \ref{lemma:perturbationlemma}, $(c)$ is obtained via Lemma \ref{lemma:xAx}, and $(d)$ is achieved via Lemma \ref{lemma:AB} by letting $\mathbf{C}=(\bm{\Sigma}^\AD_j)^{-1}\bar{\mathbf{R}}_{\mathbf{q}_j}(\bm{\Sigma}^\AD_j)^{-1}$. With all terms having been derived, we complete the proof of Theorem \ref{lemma:SEMMMSE}.

\section{Proof of Theorem \ref{lemma:SESMMSE}}
\label{app:lemmaSMMSE}

For brevity during the following derivations, we first define $\bm{\Omega}_j=\frac{1}{Mp_{j,k}}\left(\hat{\mathbf{H}}_{j,j}\mathbf{P}_{j}(\hat{\mathbf{H}}_{j,j})^H+\mathbf{Z}_j^\mathrm{S}+\sigma^2\mathbf{I}_M\right)$, $\bm{\Omega}_{j,k}=\bm{\Omega}_j-\frac{1}{M}\hat{\mathbf{h}}_{j,jk}\hat{\mathbf{h}}_{j,jk}^H$, and $\bm{\Omega}_{j,kk^\prime}=\bm{\Omega}_{j,k}-\frac{p_{j,k^\prime}}{Mp_{j,k}}\hat{\mathbf{h}}_{j,jk^\prime}\hat{\mathbf{h}}_{j,jk^\prime}^H$ for $k\neq k^\prime$. By using the definition of $\bm{\Omega}_j$, $\mathbf{v}_{j,k}^{\mathrm{M}}$ in (\ref{eq:vjkSMMSE}) can be rewritten as $\mathbf{v}_{j,k}^{\mathrm{S}}=\frac{1}{M}\bm{\Omega}_j^{-1}\hat{\mathbf{h}}_{j,jk}$.

The derivations of $A_{j,k}$, $B_{j,k}$, $C_{j,k}$, $E_{j,k}$, $F_{j,k}$, and $G_{j,k}$ related to Theorem \ref{lemma:SESMMSE} are similar to the counterparts in Theorem  \ref{lemma:SEMMMSE}. In what follows, we derive the asymptotic expression of $D_{j,k}$ and let $d_{ik^\prime}=\mathbb{E}\{|(\mathbf{v}_{j,k}^{\mathrm{S}})^H\hat{\mathbf{h}}_{j,ik^\prime}|^2\}$ for concision. Owing to the pilot contamination, two cases should be discussed.

i) When $i\neq j$ and $(i,k^\prime)\in \mathcal{P}_{j,k}$, $\hat{\mathbf{h}}_{j,jk}$ and $\hat{\mathbf{h}}_{j,ik^\prime}$ are correlated with each other. According to (\ref{eq:corrh1h2}), we have
\begin{equation}\label{eq:corrh1h2Appendix}
  \mathbb{E}\{\hat{\mathbf{h}}_{j,ik^\prime}\hat{\mathbf{h}}_{j,jk}\} =\frac{\sqrt{p_{i,k^\prime}}}{\sqrt{p_{j,k}}}\mathbf{R}_{j,i k^\prime}(\mathbf{R}_{j,jk})^{-1}\mathbf{B}_{j,jk}.
\end{equation}
In this case, it holds that
\begin{equation}\label{eq:djkS}
  \begin{split}
     &d_{ik^\prime}\\
     &= \mathbb{E}\left\{\left|\frac{\hat{\mathbf{h}}_{j,jk}^H\bm{\Omega}_j^{-1}\hat{\mathbf{h}}_{j,ik^\prime}}{M}\right|^2\right\}
\overset{(a)}{=} \mathbb{E}\left\{\left|\frac{\frac{\hat{\mathbf{h}}_{j,jk}^H\bm{\Omega}_{j,jk}^{-1}\hat{\mathbf{h}}_{j,ik^\prime}}{M}}
{1+\frac{\hat{\mathbf{h}}_{j,jk}^H\bm{\Omega}_{j,jk}^{-1}\hat{\mathbf{h}}_{j,ik^\prime}}{M}}\right|^2\right\}\\
       &\overset{(b)}{\approx} \mathbb{E}\left\{\left|\frac{\frac{\hat{\mathbf{h}}_{j,jk}^H\bm{\Omega}_{j}^{-1}\hat{\mathbf{h}}_{j,ik^\prime}}{M}}
{1+\frac{\hat{\mathbf{h}}_{j,jk}^H\bm{\Omega}_{j}^{-1}\hat{\mathbf{h}}_{j,ik^\prime}}{M}}\right|^2\right\}
\overset{(c)}{\approx} \left|\frac{\frac{\tr(\mathbb{E}\{\hat{\mathbf{h}}_{j,ik^\prime}\hat{\mathbf{h}}_{j,jk}\}\bm{\Omega}^{-1}_j)}{M}}
{1+\frac{\tr(\mathbf{B}_{j,jk}\bm{\Omega}^{-1}_j)}{M}}\right|^2\\
& \overset{(d)}{\approx}\left|\frac{\frac{\mathrm{tr}(\frac{\sqrt{p_{ik^\prime}}}{\sqrt{p_{j k}}}\mathbf{R}_{j,i k^\prime}(\mathbf{R}_{j,jk})^{-1}\mathbf{B}_{j,jk}\bm{\Gamma}_{j,jk})}{M}}{1+\frac{\mathrm{tr}(\mathbf{B}_{j,jk}\bm{\Gamma}_{j,jk})}{M}}\right|^2.
  \end{split}
\end{equation}
Note that $(a)$ follows Lemma \ref{lemma:MP}, $(b)$ utilizes Lemma \ref{lemma:perturbationlemma}, $(c)$ is obtained via Lemma \ref{lemma:xAx}, and $(d)$ is achieved via Lemma \ref{lemma:AB} by letting $L=1$, $\mathbf{A}=\mathbf{B}_{j,jk}$, $\bm{\Delta}_{jk^\prime}=\frac{p_{j,k^\prime}}{p_{j,k}}\mathbf{B}_{j,jk^\prime}$, $\mathbf{D}=\frac{\mathbf{Z}^{\mathrm{S}}_{j}}{p_{j,k}M}$, and $\alpha=\frac{\sigma^2}{p_{j,k}M}$.

ii) When $i\neq j$ and $(i,k^\prime)\notin \mathcal{P}_{j,k}$, $\hat{\mathbf{h}}_{j,jk}$ and $\hat{\mathbf{h}}_{j,ik^\prime}$ are statistically uncorrelated. In this case, it follows that
\begin{equation}\label{eq:djkS2}
  \begin{split}
     d_{ik^\prime}&= \frac{1}{M^2} \mathbb{E}\left\{\hat{\mathbf{h}}_{j,jk}^H\bm{\Omega}_j^{-1}\hat{\mathbf{h}}_{j,ik^\prime}\hat{\mathbf{h}}_{j,ik^\prime}^H\bm{\Omega}_j^{-1}\hat{\mathbf{h}}_{j,jk}\right\}\\
&\overset{(a)}{=} \frac{1}{M^2}\mathbb{E}\left\{\frac{\hat{\mathbf{h}}_{j,jk}^H\bm{\Omega}_{j,jk}^{-1}\hat{\mathbf{h}}_{j,ik^\prime}\hat{\mathbf{h}}_{j,ik^\prime}^H\bm{\Omega}_{j,jk}^{-1}\hat{\mathbf{h}}_{j,jk}}
{|1+\frac{1}{M}\hat{\mathbf{h}}_{j,jk}^H\bm{\Omega}_{j,jk}^{-1}\hat{\mathbf{h}}_{j,jk}|^2}\right\}\\
&\overset{(b)}{\approx}\frac{1}{M^2}\mathbb{E}\left\{\frac{\hat{\mathbf{h}}_{j,jk}^H\bm{\Omega}_{j}^{-1}\hat{\mathbf{h}}_{j,ik^\prime}\hat{\mathbf{h}}_{j,ik^\prime}^H\bm{\Omega}_{j}^{-1}\hat{\mathbf{h}}_{j,jk}}
{|1+\frac{1}{M}\hat{\mathbf{h}}_{j,jk}^H\bm{\Omega}_{j}^{-1}\hat{\mathbf{h}}_{j,jk}|^2}\right\}\\
&\overset{(c)}{\approx}\frac{1}{M^2}\mathbb{E}\left\{\frac{\tr(\mathbf{B}_{j,jk}\bm{\Omega}_{j}^{-1}\mathbf{B}_{j,ik^\prime}\bm{\Omega}_{j}^{-1})}
{|1+\frac{1}{M}\tr(\mathbf{B}_{j,jk}\bm{\Omega}_{j}^{-1})|^2}\right\}\\
&\overset{(d)}{\approx}\frac{\frac{\mathrm{tr}(\mathbf{B}_{j,jk}\tilde{\bm{\Gamma}}^{\prime}_{j,ik^\prime})}{M^2}}{\left|1+\frac{\mathrm{tr}(\mathbf{B}_{j,jk}\bm{\Gamma}_{j,jk})}{M}\right|^2}.
\end{split}
\end{equation}
Note that $(a)$ follows Lemma \ref{lemma:MP}, $(b)$ utilizes Lemma \ref{lemma:perturbationlemma}, $(c)$ is obtained via Lemma \ref{lemma:xAx}, and $(d)$ is achieved via Lemma \ref{lemma:AB} by letting $\mathbf{C}=\mathbf{B}_{j,ik^\prime}$. Due to $D_{j,k}=\sum_{i\neq j}^{K}\sum_{k^\prime=1}^{K} p_{i,k^\prime}d_{ik^\prime}$, we can obtain (\ref{eq:DjkSMMSE}) following (\ref{eq:djkS}) and (\ref{eq:djkS2}).  With all terms having been derived, we complete the proof of Theorem \ref{lemma:SESMMSE}.

%\section{Proof of}
%\label{app:lemma2}

%\section*{Acknowledgment}
%
%
%The authors would like to thank the Editor and
%Reviewers for reviewing the manuscript.

\ifCLASSOPTIONcaptionsoff
  \newpage
\fi



%\bibliographystyle{IEEEtran}

\bibliographystyle{IEEEtran}

\bibliography{IEEEabrv,mybibfileBiGAMPADCmmWave,mybibfileMaMIMOADC}


%\end{IEEEbiography}


%\newpage



%\vfill


\end{document}